\documentclass{article}
\usepackage[utf8]{inputenc}
\usepackage{amsmath,amssymb,amsfonts,stmaryrd}
\usepackage{physics}
\usepackage{dsfont}
\usepackage{bbm}
\usepackage{graphicx}
\usepackage{xcolor}
\usepackage{graphicx}
\usepackage{cancel}
\usepackage[numbers,sort&compress,square]{natbib}
\usepackage[linktocpage]{hyperref}
\usepackage{mathtools}
\usepackage{bm} 
\usepackage{subfigure} 
\usepackage{environ}
\usepackage{url}
\usepackage{comment}
\usepackage{authblk}
\usepackage[margin=0.75in]{geometry}
\usepackage[mathscr]{euscript}
\usepackage[scr=boondox]{mathalpha}
\hypersetup{colorlinks=true,citecolor=blue,linkcolor=blue, urlcolor=blue, breaklinks=true}

\usepackage{amsthm}

\newtheorem{theorem}{Theorem}[section]
\newtheorem{lemma}{Lemma}[section]

\NewEnviron{eqs}{%
\begin{align}\begin{split}
    \BODY
\end{split}\end{align}
}

\title{Crystalline Equivalence Principle for Anomalies}
\author[1,2]{Ching-Yu Yao}
\author[3]{Ryohei Kobayashi}
\affil[1]{\textit{Institute for Solid State Physics, University of Tokyo, Chiba 277-8583, Japan}}
\affil[2]{\textit{Department of Physics, University of Tokyo, Tokyo 113-0033, Japan}}
\affil[3]{\textit{Department of Applied Physics, University of Tokyo, Tokyo 113-8656, Japan}}

\date{}
\begin{document}
\maketitle

\pagenumbering{arabic}
\setcounter{tocdepth}{1}
\setcounter{secnumdepth}{2}

\begin{abstract}
We develop a microscopic characterization of crystalline symmetry anomalies in bosonic lattice systems in (1+1) and (2+1) dimensions with a combination of crystalline and finite unitary internal symmetry. We construct anomaly indices of symmetry operators and derive their cocycle conditions and coboundary ambiguities. For generalized crystalline symmetries combining permutations of onsite Hilbert spaces with quantum cellular automata, anomaly indices take values in $H^3(B\widetilde \Gamma,U(1)_r)$ and $H^4(B\widetilde \Gamma,U(1)_r)$ respectively, where $\widetilde \Gamma$ is the total group of crystalline and internal symmetries, and the coefficient twist $r$ accounts for spatial orientation reversal. When crystalline symmetries act solely by permuting onsite Hilbert spaces, the anomaly classes restrict trivially to the spatial subgroup. These results establish a crystalline equivalence principle for bosonic anomalies at the microscopic level, without assuming a continuum description or emergent Lorentz invariance. We prove that a nontrivial total anomaly index obstructs any symmetric invertible phase. The framework incorporates modulated symmetries where spatial symmetries permute internal symmetries. Our results reproduce LSM and SPT--LSM constraints, with the latter illustrating how spectral-sequence differentials can trivialize candidate LSM anomaly indices.
\end{abstract}
\tableofcontents
\newpage
\newpage
\section{Introduction}\label{intro:sec}
Crystalline symmetries place fundamental constraints on the phases and low-energy dynamics of quantum many-body systems. A basic example is the Lieb--Schultz--Mattis (LSM) theorem: a spin chain with translation and internal symmetries cannot host a trivial symmetric gapped state when half-odd integer spin is located per unit cell~\cite{LiebSchultzMattis1961}. Its higher-dimensional extensions reveal a general connection between lattice symmetry actions and the necessity of non-trivial low energy spectrum~\cite{Oshikawa2000,Hastings2004}. Subsequent developments have established such constraints involving generic crystalline symmetries, and generic distribution of projective representations at onsite Hilbert spaces~\cite{WatanabePoVishwanathZaletel2015,PoWatanabeJianZaletel2017}. These development lead to various complementary ways to formulate LSM theorems for more general combination of crystalline and internal symmetries, using the method of e.g., lattice homotopy and operator algebraic approach~\cite{ElseThorngren2020,OgataTasaki2019,OgataTachikawaTasaki2021,YaoFurusaki2022}.

The interpretation of LSM constraints as symmetry anomalies has been particularly fruitful. Connections between the LSM constraints and the symmetry action on the anomalous surface of symmetry-protected-topological (SPT) phase has been discussed in Ref.~\cite{ChengZaletelBarkeshliVishwanathBonderson2016}. 
The interpretation of LSM constraints in terms of continuum anomalies has been developed in Refs.~\cite{ChoHsiehRyu2017,MetlitskiThorngren2018,YeGuoHeWangZou2022,ChengSeiberg2023}. In this description, microscopic lattice translation by a lattice unit acts in the infrared as  an internal symmetry, which is called an {emanant symmetry} \cite{ChengSeiberg2023}. The corresponding LSM constraint is then encoded in a mixed 't Hooft anomaly between this emanant symmetry and other internal symmetries inherited from that of the UV lattice model. The dynamical consequences of LSM type constraints involving generalized symmetries, including higher-form or non-invertible symmetries, have been discussed in e.g.,~\cite{Kobayashi2019lsm,SeibergSeifnashriShao2024,SeibergShaoZhang2025, Pace2025SPTLSM}. Throughout these developments, a central issue is to identify precisely which information in a microscopic symmetry action produces an anomaly and what kind of dynamical consequences follow from it.

In parallel, crystalline symmetry has become an essential ingredient in the classification of topological phases. For free fermions, topological crystalline insulators and superconductors have been studied using band topology, $K$-theory, symmetry indicators, and topological quantum chemistry~\cite{Fu2011,ShiozakiSato2014,ShiozakiSatoGomi2016,ShiozakiSatoGomi2017,PoVishwanathWatanabe2017,BradlynEtAl2017}. Interacting crystalline phases admit bottom-up real-space constructions in terms of lower-dimensional building blocks and their junctions~\cite{SongHuangFuHermele2017,HuangSongHuangHermele2017,ElseThorngren2019,SongFangQi2020}. Generalized homology and the Atiyah--Hirzebruch spectral sequence organize the consistency conditions and equivalences among such constructions~\cite{ShiozakiXiongGomi2023}. Explicit constructions of crystalline phases have also tested and refined the relation between the classification of SPT phases with internal and crystalline symmetries~\cite{ChengWang2022,ZhangWangYangQiGu2020,ZhangYangQiGu2022,ZhangNingQiGu2025}.

A unifying principle is the crystalline equivalence principle formulated by Thorngren and Else~\cite{ThorngrenElse2018}. For crystalline topological phases on Euclidean space, it identifies the classification of crystalline phases with that for the same abstract symmetry group acting internally. Mathematical formulations of the crystalline equivalence principle have been developed within generalized-homology descriptions of invertible phases~\cite{Debray2021}. The equivalence principle is also supported by microscopic probes of crystalline phases such as partial rotation operations~\cite{ShiozakiShapourianRyu2017, ManjunathBarkeshli2021,ZhangManjunathKobayashiBarkeshli2023,ManjunathCalveraBarkeshli2024}.

These results motivate an analogous question for crystalline anomalies. If an infrared theory has the Lorentz invariance, a crystalline operation is expected to act in IR through a combination of a spacetime Lorentz transformation and an internal transformation, making an internal-symmetry description natural. However, a general microscopic lattice system need not possess an IR limit with Lorentz symmetry. Moreover, it remains unclear if general crystalline anomalies are described through the inflow from bulk crystalline SPT phases, or if the classification of crystalline anomalies coincides with that of crystalline SPTs in the bulk. Establishing the classification of crystalline anomalies therefore requires a direct microscopic characterization and a systematic microscopic analysis of its dynamical consequences.

There has recently been substantial progress of characterizing the 't Hooft anomalies at microscopic lattice models~\cite{ElseNayak2014, KawagoeLevin2021,Seifnashri2024,KapustinSopenko2025,KawagoeShirley2025, kobayashi2025generalizedstatistics, TuLongElse2026,CzajkaGeikoThorngren2025,FengKobayashiChenRyu2026, kapustin2025anomaly2d}. In particular, microscopic anomalies involving translation have already been studied in quantum spin chains~\cite{Seifnashri2024,KapustinSopenko2025}.  These advances enable a direct microscopic characterization of crystalline anomalies in a tensor product Hilbert space, without relying on a continuum description or anomaly inflow from bulk crystalline SPT phases.

In this work, we develop such a microscopic construction for bosonic systems in (1+1) and (2+1) dimensions with a combination of crystalline and finite unitary internal symmetry. We obtain explicit cocycle data for combined internal and crystalline symmetry actions, identify their ambiguities to define cohomological data, and establish a dynamical consequence as an obstruction to symmetric invertible states.

\subsection{Summary of results}
\label{intro:summary}

We consider a bosonic tensor product Hilbert space
\begin{equation}
  \mathcal H=\bigotimes_{x\in\Lambda}\mathcal H_x
\end{equation}
on a lattice $\Lambda$ without boundary in spatial dimension $d=1,2$, where each onsite Hilbert space is assumed to be isomorphic for simplicity. The internal symmetry group $G$ is finite and unitary, and its operators $U(g)$ preserve locality. 
A spatial symmetry group $G_{\mathrm{sp}}$ acts on the lattice and may also act on the internal symmetry group $G$ by conjugation, through a homomorphism $\phi:G_{\mathrm{sp}}\to\operatorname{Aut}(G)$. When this action is nontrivial, the internal symmetry is referred to as a {modulated symmetry}~\cite{Bulmash2025,PaceBulmash2026}. 

Write the total symmetry group as
\begin{equation}
  \Gamma=G\rtimes_\phi G_{\mathrm{sp}},
  \qquad r:G_{\mathrm{sp}}\longrightarrow\{+1,-1\}
\end{equation}
where $r$ characterizes the spatial symmetry action that reverses the orientation of the space. We denote by $U(1)_r$ the module on which $G$ acts trivially and $s\in G_{\mathrm{sp}}$ acts as $c\mapsto c^{r(s)}\in U(1)$. 

First, suppose that the spatial action consists only of permutations of local Hilbert spaces,
\begin{equation}
  V_0(s)\bigotimes_{x\in\Lambda}|\phi_x\rangle_x
  =\bigotimes_{x\in\Lambda}|\phi_x\rangle_{sx}.
  \label{intro:permutation}
\end{equation}
For a given symmetry operator, we construct a microscopic anomaly index valued in
\begin{equation}
  \ker\!\left[
    H^{d+2}\!\left(B\Gamma,U(1)_r\right)
    \xrightarrow{\;\pi_*\;}
    H^{d+2}\!\left(BG_{\mathrm{sp}},U(1)_r\right)
  \right],
  \qquad d=1,2,
  \label{intro:kernel}
\end{equation}
where $\pi(s)=(1,s)$. The kernel expresses the fact that the pure permutation action carries no 't Hooft anomaly: every anomaly detected in this setting restricts trivially to $G_{\mathrm{sp}}$. 

The explicit construction of the anomaly index gives the cocycles directly from symmetry operators. In one spatial dimension, the above anomaly index consists of the components
\begin{equation}
  \omega^{0,3},\quad\omega^{1,2},\quad\omega^{2,1}.
\end{equation}
In two spatial dimensions, the index consists of the components labeled by
\begin{equation}
  \omega^{0,4},\quad\omega^{1,3},\quad
  \omega^{2,2},\quad\omega^{3,1}.
\end{equation}
These anomaly indices $\{\omega^{p,q}\}$ are valued in
\begin{equation}
  C^{p,q}=C^p\!\left(BG_{\mathrm{sp}},C^q(BG,U(1))\right),
  \label{intro:double}
\end{equation}
and obey the cocycle conditions of the double complex with the horizontal and vertical differentials $d_h, d_v$ raising the degrees of $p,q$ respectively. 
Possible ambiguities of the anomaly indices give the coboundary transformations again defined in the double complex. The collection of the anomaly indices $\omega=\sum_{p+q=d+2}\omega^{p,q}$ up to possible ambiguities gives a cohomology of the total complex, which is precisely given by Eq.~\eqref{intro:kernel}. In particular, the each component of the anomaly index $\omega^{p,q}$ is a part of a single total cocycle $\omega$ and organized into a single cohomology invariant.

Second, we allow generalized spatial symmetry operators
\begin{equation}
  V(s)=U(s)V_0(s),
  \label{intro:generalized}
\end{equation}
where $U(s)$ is a locality preserving unitary, i.e., quantum cellular automata (QCA). Multiplication of crystalline symmetries may then produce an internal symmetry,
\begin{equation}
  V(s_1)V(s_2)=U\!\left(e(s_1,s_2)\right)V(s_1s_2),
\end{equation}
where we assume $e(s_1,s_2)\in Z(G)$. These relations define an extension
\begin{equation}
  1\longrightarrow G\longrightarrow\widetilde\Gamma
  \xrightarrow{\;p\;}G_{\mathrm{sp}}\longrightarrow1.
  \label{intro:extension}
\end{equation}
For the extensions treated here, the anomaly index takes values in
\begin{equation}
  [\omega]\in H^{d+2}\!\left(B\widetilde\Gamma,U(1)_r\right).
  \label{intro:full}
\end{equation}
Here the orientation action on the coefficients is pulled back along $p$. When the extension splits, the above formula gives the full group cohomology of the semidirect product. We note that the restriction to the spatial subgroup generally no longer vanishes, since the spatial action can itself contribute to the anomaly data. For a non-split extension, there is generally no spatial subgroup inclusion, accordingly the formula~\eqref{intro:kernel} is replaced by~\eqref{intro:full}.

We establish the dynamical consequence of the anomaly index by showing that a symmetric invertible state must have a trivial anomaly class. Therefore, $[\omega]\ne0$ rules out any symmetry-preserving invertible phase. This conclusion concerns the combined symmetry action. It does not require the internal symmetry action to be non-onsite; indeed, an onsite internal symmetry can participate in a mixed LSM anomaly. 

Equations~\eqref{intro:kernel} and~\eqref{intro:full} establish a microscopic form of crystalline equivalence for the bosonic anomalies. The full index has the same cohomological description as the corresponding bosonic continuum anomaly of internal symmetry~\cite{ChenGuLiuWen2013}. This statement is obtained purely from lattice symmetry operators without assuming specific properties of the IR theories. 

\subsection{Relation to previous work}
\label{intro:previous}

\paragraph{Microscopic lattice anomalies}

Our strategy to obtain anomaly indices builds on Else--Nayak approach~\cite{ElseNayak2014} and its two-dimensional extension~\cite{KawagoeShirley2025}. It should also be compared with the microscopic formula of anomalies in (1+1)D translationally symmetric spin chains~\cite{Seifnashri2024, KapustinSopenko2025}. For one-dimensional translation symmetry commuting with $G$, the anomaly indices are valued in $H^3(BG,U(1))$ and $H^2(BG,U(1))$, where the latter $H^2$ index labels the LSM anomaly. Our microscopic characterization provides explicit formula for each component of anomalies $\omega^{p,q}$ in both one and two spatial dimensions, generalizing the above microscopic formula for LSM anomalies.
Our setup incorporates the modulated symmetry where the crystalline symmetry acts on the internal symmetry by an automorphism and admits orientation reversing spatial symmetries such as reflection.

\paragraph{Double complexes and spectral sequences}
A topological theory of LSM constraints has been developed in Ref.~\cite{ElseThorngren2019} using equivariant homology and double complexes, and discussed the role of the Lyndon--Hochschild--Serre (LHS) spectral sequence. Ref.~\cite{WangNingCheng2021} gave explicit cochain descriptions of the LHS spectral sequence and interpreted its differentials through domain-wall decoration. These developments are closely related to the real-space spectral-sequence approach to crystalline phases~\cite{ShiozakiXiongGomi2023}. 
Ref.~\cite{ning2026matrixproductstatesmodulated} developed the description of the LSM type theorems using the LHS spectral sequence in (1+1)D assuming the MPS representation of the ground state.
More recently, Ref.~\cite{ZhangLinYangWang2026} studied $E_\infty^{1,2}$-type translation anomalies and their consequences for deconfined criticality.

The use of a double complex and spectral sequence for crystalline anomalies therefore precedes the present work. Our contribution is a microscopic characterization of each cochain data of anomaly indices entering the LHS spectral sequence purely based on lattice symmetry operators, together with the proof of their obstruction to symmetric invertible states. For Eq.~\eqref{intro:extension}, the relevant spectral sequence is
\begin{equation}
  E_2^{p,q}
  =H^p\!\left(BG_{\mathrm{sp}},H^q(BG,U(1))\right)
  \ \Longrightarrow\
  H^{p+q}\!\left(B\widetilde\Gamma,U(1)_r\right).
  \label{intro:LHS}
\end{equation}
The microscopic components $\omega^{p,q}$ are cochains before taking cohomology. The $E_2$ page organizes candidate layers, while differentials impose compatibility and identify trivializations. As described below, such differentials may trivialize some of the LSM anomalies encoded in $\omega^{p,q}$ that leads to the mechanism known as the SPT--LSM theorem.

\paragraph{SPT--LSM theorems and magnetic translations}
An SPT--LSM theorem permits a symmetric invertible ground state compatible with the seemingly nontrivial LSM anomaly index. Such phenomena were established for magnetic translations and related settings in Refs.~\cite{Lu2024,YangJiangVishwanathRan2018,LuRanOshikawa2020,JiangChengQiLu2021}. Their topological and spectral-sequence interpretations were discussed in Refs.~\cite{ElseThorngren2020,WangNingCheng2021}. In this situation, a nonzero candidate LSM layer does not necessarily give a nonzero total anomaly, because a differential in the LHS spectral sequence can make that layer exact.

For instance, an example discussed in \cite{YangJiangVishwanathRan2018} makes this distinction explicit at the microscopic level. Suppose that a magnetic translation symmetry of a square lattice satisfies the algebra
\begin{equation}
  T_xT_yT_x^{-1}T_y^{-1}=U(a),\qquad a\in Z(G),
\end{equation}
and translations commute with the internal symmetry. A projective representation $\alpha\in H^2(BG,U(1))$ per unit cell defines the candidate layer $u\otimes\alpha\in E_2^{2,2}$ that corresponds to the LSM anomaly, where $u$ generates $H^2(B\mathbb Z^2,\mathbb Z)$. The differential
\begin{equation}
  d_2^{0,3}:H^3(BG,U(1))
  \longrightarrow H^2\!\left(B\mathbb Z^2,H^2(BG,U(1))\right)
\end{equation}
is determined by the flux $a$ inserted at each plaquette. Using the microscopic formula of anomaly indices, we show that a symmetric invertible state requires a class $[\lambda]\in H^3(BG,U(1))$ satisfying
\begin{equation}
  \alpha=[\delta_a^3\lambda],
  \label{intro:dyonic}
\end{equation}
where $\delta_a^3$ is the slant product operation that corresponds to twisted compactification of the SPT phase with $a$ flux insertion. Thus the obstruction to symmetric invertible states from this layer lies in $H^2(BG,U(1))/\operatorname{im}\delta_a^3$. Eq.~\eqref{intro:dyonic} recovers the anomaly-cancellation condition underlying the dyonic SPT--LSM theorem formulated in Ref.~\cite{YangJiangVishwanathRan2018}. 

\paragraph{Modulated symmetries}
Modulated symmetries provide a setting in which spatial and internal symmetries do not commute. LSM constraints for dipole-conserving and fractonic systems were studied in Refs.~\cite{HeYouPrem2020,DubinkinMayMannHughes2021}. Modulated SPT phases, their matrix product state descriptions, anomaly inflow, and gauging have been developed in e.g.,~\cite{HanLakeLamVerresenYou2024,Lam2024,EbisuHondaNakanishi2024,PaceDelfinoLamAksoy2025,Aksoy2024LSM,Pace2026spacetimeenrichedSymTFT,Bulmash2025,SaitoCaoHanEbisu2025,yao2025latticetranslationmodulatedsymmetries}. Ref.~\cite{PaceBulmash2026} formulated stratified anomalies and derived LSM and SPT--LSM constraints for modulated symmetries. Their Sec.~3.1 also decomposes a crystalline symmetry operator into a locality-preserving internal part and a permutation of local Hilbert spaces, closely matching our starting point. Our setup includes the nontrivial automorphism action that characterizes finite modulated symmetries. Our contribution is to establish a microscopic formula each layer of anomalies for modulated symmetries, which organizes into a group cohomology of a total symmetry group where crystalline symmetries act on internal symmetries by automorphisms.

\subsection{Organization of the paper}
The remainder of this paper is organized as follows. Sec.~\ref{Sec2} develops the microscopic anomaly indices in (1+1) dimensions and illustrates their relation to the LSM anomaly. Sec.~\ref{Sec3} extends the construction to (2+1) dimensions. Sec.~\ref{Sec4} considers generalized spatial symmetries that combine lattice permutations with QCAs and applies the framework to magnetic translations and SPT--LSM constraints. Technical details of this work is relegated to appendices.
\section{Anomalies in 1+1D}\label{Sec2}
In this section, we consider a 1D lattice $\Lambda$ without boundary and a tensor product Hilbert space $\mathcal{H}=\bigotimes_{x\in\Lambda} \mathcal{H}_x$, where the onsite Hilbert spaces $\mathcal{H}_x$ are all isomorphic and the system contains no fermionic degrees of freedom.

A finite locality-preserving unitary symmetry is described by a representation
\begin{align}
    \begin{split}
        U:G\to&\mathrm{QCA}(\mathcal{H}),\\
        g\mapsto& U(g)
    \end{split}
\end{align}
where $G$ is the finite internal symmetry group and $\mathrm{QCA}(\mathcal{H})\subseteq\mathrm{End}(\mathcal{H})$ denotes the group of QCAs acting on $\Lambda$. Since $G$ is finite, every QCA in the image of $U$ is a finite-depth quantum circuit (FDQC).

On the other hand, a spatial symmetry is specified by a faithful group action $G_{\mathrm{sp}}\curvearrowright\Lambda$. This action induces a unitary operator $V(s)$ for each $s\in G_{\mathrm{sp}}$, defined by
\begin{equation}
    V(s)\bigotimes_{x\in\Lambda}|\phi_x\rangle_x:=\bigotimes_{x\in\Lambda}|\phi_x\rangle_{sx}.
\end{equation}
Thus, the spatial symmetry considered here acts solely by permuting the local Hilbert spaces. In particular, $V(s_1s_2)=V(s_1)V(s_2)$ for every $s_1,s_2\in G_{\mathrm{sp}}$.

Since $V(s)^{-1}U(g)V(s)$ is an internal symmetry operator for each $g\in G$ and $s\in G_{\mathrm{sp}}$, there exists a group homomorphism
\begin{equation}
    \phi:G_{\mathrm{sp}}\to\mathrm{Aut}(G),
\end{equation}
such that 
\begin{equation}
    U(^sg)=V(s)^{-1}U(g)V(s),\quad {}^sg:=\phi(s)^{-1}(g).
\end{equation}
In particular, this convention gives ${}^{s_1s_2}g={}^{s_2}({}^{s_1}g)$. The full symmetry group is therefore described by the semidirect product $G\rtimes_\phi G_{\mathrm{sp}}$.

In this section, we characterize the ‘t Hooft anomaly of this symmetry using an Else-Nayak-type characterization \cite{ElseNayak2014}. Detailed calculations are provided in Appendix \ref{Conditions}. We further show that, although a nontrivial ‘t Hooft anomaly does not necessarily obstruct onsite-ability of the symmetry, it does obstruct the existence of a symmetric invertible state.

\subsection{‘t Hooft anomaly}
Let $I\subset\Lambda$ be a sufficiently large connected segment. For each $g\in G$, we choose a truncation $U_I(g)$ of the symmetry operator $U(g)$ such that, for every two segments $I$ and $I'$ sharing the same right (left) endpoint, the actions of $U_I(g)$ and $U_{I'}(g)$ coincide in a neighborhood of that endpoint.

Since $U(g_1)U(g_2)=U(g_1g_2)$, the operator $U_I(g_1)U_I(g_2)U_I(g_1g_2)^{-1}$ is supported near the two endpoints of $I$. We therefore define
\begin{equation}
    U_I(g_1)U_I(g_2)=\Omega_{\partial I}^{0,2}(g_1,g_2)U_I(g_1g_2),
\end{equation}
where $\Omega_{\partial I}^{0,2}(g_1,g_2)=\Omega_{I^+}^{0,2}(g_1,g_2)\Omega_{I^-}^{0,2}(g_1,g_2)$, with $\Omega_{I^+}^{0,2}(g_1,g_2)$ and $\Omega_{I^-}^{0,2}(g_1,g_2)$ supported near the right and left endpoints of $I$, respectively. We choose these endpoint operators consistently, so that $\Omega_{I^+}^{0,2}(g_1,g_2)$ ($\Omega_{I^-}^{0,2}(g_1,g_2)$) is the same for any two segments sharing the same right (left) endpoint.

Similarly, since $V(s)U(^sg)V(s)^{-1}=U(g)$ and $V(s)$ maps $x\in\Lambda$ to $sx\in\Lambda$, $V(s)U_{s^{-1}I}(^sg)V(s)^{-1}U_I(g)^{-1}$ is supported near the two endpoints of $I$. We therefore define
\begin{equation}
    V(s)U_{s^{-1}I}(^sg)=\Omega_{\partial I}^{1,1}(s;g)U_I(g)V(s).
\end{equation}
where $\Omega_{\partial I}^{1,1}(s;g)=\Omega_{I^+}^{1,1}(s;g)\Omega_{I^-}^{1,1}(s;g)$, with $\Omega_{I^+}^{1,1}(s;g)$ and $\Omega_{I^-}^{1,1}(s;g)$ supported near the right and left endpoints of $I$, respectively. Again, we choose these endpoint operators consistently for segments sharing the same endpoint.

The consistency conditions for these endpoint operators can be obtained by evaluating the following expressions in two different ways:
\begin{align}
    &U_I(g_1)U_I(g_2)U_I(g_3),\\
    &V(s)U_{s^{-1}I}({}^sg_1)U_{s^{-1}I}({}^sg_2),\\
    &V(s_1)V(s_2)U_{(s_1s_2)^{-1}I}({}^{s_1s_2}g).
\end{align}
This gives
\begin{align}
    &\Omega_{\partial I}^{0,2}(g_1,g_2)\Omega_{\partial I}^{0,2}(g_1g_2,g_3)={}^{g_1}\Omega_{\partial I}^{0,2}(g_2,g_3)\Omega_{\partial I}^{0,2}(g_1,g_2g_3),\label{1D (0,3)}\\
    &\Omega_{\partial I}^{1,1}(s;g_1){}^{g_1}\Omega_{\partial I}^{1,1}(s;g_2)\Omega_{\partial I}^{0,2}(g_1,g_2)={}^s\Omega_{\partial(s^{-1}I)}^{0,2}(^sg_1,^sg_2)\Omega_{\partial I}^{1,1}(s;g_1g_2),\label{1D (1,2)}\\
    &\Omega_{\partial I}^{1,1}(s_1s_2;g)={}^{s_1}\Omega_{\partial(s_1^{-1}I)}^{1,1}(s_2;{}^{s_1}g)\Omega_{\partial I}^{1,1}(s_1;g).\label{1D (2,1)}
\end{align}
Here we use the notation
\begin{align}
    &{}^g\mathcal{O}:=U_I(g)\mathcal{O}U_I(g)^{-1},\\
    &{}^s\mathcal{O}:=V(s)\mathcal{O}V(s)^{-1}.
\end{align}

A spatial symmetry may reverse the orientation of the lattice. If $V(s)$ reverses the orientation, then ${}^s\Omega_{(s^{-1}I)^+}$ is actually supported at the left endpoint of $I$, rather than the right endpoint. Let $r:G_{\mathrm{sp}}\to \{\pm1\}$ denote the orientation character, with $r(s)=1$ for orientation-preserving symmetries and $r(s)=-1$ for orientation-reversing symmetries. Accordingly, we define the $G_{\mathrm{sp}}$-action on $U(1)$ by $c\mapsto{}^sc:=c^{r(s)}$ for every $c\in U(1)$ and $s\in G_{\mathrm{sp}}$.

With this notation, the consistency conditions give rise to the following $U(1)$-valued phases
\begin{align}
    &\omega_{I^\pm}^{0,3}(g_1,g_2,g_3):=\Omega_{I^\pm}^{0,2}(g_1g_2,g_3)^{-1}\Omega_{I^\pm}^{0,2}(g_1,g_2)^{-1}{}^{g_1}\Omega_{I^\pm}^{0,2}(g_2,g_3)\Omega_{I^\pm}^{0,2}(g_1,g_2g_3),\\
    &\omega_{I^\pm}^{1,2}(s;g_1,g_2):=\Omega_{I^\pm}^{0,2}(g_1,g_2)^{-1}{}^{g_1}\Omega_{I^\pm}^{1,1}(s;g_2)^{-1}\Omega_{I^\pm}^{1,1}(s;g_1)^{-1}{}^s\Omega_{(s^{-1}I)^{\pm r(s)}}^{0,2}({}^sg_1,{}^sg_2)\Omega_{I^\pm}^{1,1}(s;g_1g_2),\\
    &\omega_{I^\pm}^{2,1}(s_1,s_2;g):=\Omega_{I^\pm}^{1,1}(s_1s_2;g)^{-1}{}^{s_1}\Omega_{(s_1^{-1}I)^{\pm r(s_1)}}^{1,1}(s_2;^{s_1}g)\Omega_{I^\pm}^{1,1}(s_1;g).
\end{align}
By construction, $\omega_{I^+}$ depends only on the location of the right endpoint of $I$, while $\omega_{I^-}$ depends only on the location of its left endpoint. Moreover, by definition, $\omega_{I^-}=\omega_{I^+}^{-1}$. It follows that the resulting phase is independent of the choice of segment $I$. We denote it by $\omega:=\omega_{I^+}$. These $\omega$'s are the anomaly indices. In particular, the construction of $\omega^{0,3}$ is exactly the Else-Nayak index.

There are three types of ambiguities in this construction. First, we may redefine each onsite Hilbert space by multiplying an on-site unitary map $U$. Hence, the new internal symmetry operators and spatial symmetry operators have the forms $UU(g)U^{-1}$ and $UV(s)U^{-1}$, respectively. We can therefore choose a canonical truncation of each $UU(g)U^{-1}$ to be $UU_I(g)U^{-1}$ and obtain the exact same indices.

Second, the truncation $U_I$ is not unique and can be modified by operators supported near the two endpoints of $I$. Since the modifications at the two endpoints can be made independently, the resulting $\omega$'s are invariant under changes of truncation.

Lastly, we may rescale the endpoint operators by $U(1)$-valued functions:
\begin{align}
    &\Omega_{I^\pm}^{0,2}(g_1,g_2)\mapsto\alpha^{0,2}(g_1,g_2)^{\pm1}\Omega_{I^\pm}^{0,2}(g_1,g_2),\\
    &\Omega_{I^\pm}^{1,1}(s;g)\mapsto\alpha^{1,1}(s;g)^{\pm1}\Omega_{I^\pm}^{1,1}(s;g).
\end{align}
Owing to our consistency convention for the endpoint operators $\Omega_{I^\pm}$'s, the functions $\alpha$'s are independent of $I$. Under these redefinitions, the anomaly indices transform as follows:
\begin{align}
    &\omega^{0,3}(g_1,g_2,g_3)\mapsto\omega^{0,3}(g_1,g_2,g_3)\alpha^{0,2}(g_1g_2,g_3)^{-1}\alpha^{0,2}(g_1,g_2)^{-1}\alpha^{0,2}(g_2,g_3)\alpha^{0,2}(g_1,g_2g_3),\\
    &\omega^{1,2}(s;g_1,g_2)\mapsto\omega^{1,2}(s;g_1,g_2)\alpha^{0,2}(g_1,g_2)^{-1}\alpha^{1,1}(s;g_2)^{-1}\alpha^{1,1}(s;g_1)^{-1}{}^s\alpha^{0,2}(^sg_1,^sg_2)\alpha^{1,1}(s;g_1g_2),\\
    &\omega^{2,1}(s_1,s_2;g)\mapsto\omega^{2,1}(s_1,s_2;g)\alpha^{1,1}(s_1s_2;g)^{-1}{}^{s_1}\alpha^{1,1}(s_2;{}^{s_1}g)\alpha^{1,1}(s_1;g).
\end{align}

These relations admit a natural formulation in terms of a double complex. Consider
\begin{equation}
    C^{p,q}:=C^p(BG_{\mathrm{sp}},C^q(BG,U(1))),
\end{equation}
with horizontal and vertical differentials $d_h$ and $d_v$, respectively (see Appendix \ref{Cohomology of Semidirect Products}). The $\omega^{p,q}$'s and $\alpha^{p,q}$'s naturally define elements of $C^{p,q}$. The above ambiguities can then be expressed as
\begin{align}
    &\omega^{0,3}\mapsto\omega^{0,3}+d_v\alpha^{0,2},\label{1D (0,3) ambiguity}\\
    &\omega^{1,2}\mapsto\omega^{1,2}+d_h\alpha^{0,2}-d_v\alpha^{1,1},\label{1D (1,2) ambiguity}\\
    &\omega^{2,1}\mapsto\omega^{2,1}+d_h\alpha^{1,1}.\label{1D (2,1) ambiguity}
\end{align}

Finally, by evaluating the following expressions in two different ways:
\begin{align}
    &\Omega_{I^-}^{0,2}(g_1,g_2)\Omega_{I^-}^{0,2}(g_1g_2,g_3)\Omega_{I^-}^{0,2}(g_1g_2g_3,g_4),\\
    &\Omega_{I^-}^{1,1}(s;g_1){}^{g_1}\Omega_{I^-}^{1,1}(s;g_2)\Omega_{I^-}^{0,2}(g_1,g_2){}^{g_1g_2}\Omega_{I^-}^{1,1}(s;g_3)\Omega_{I^-}^{0,2}(g_1g_2,g_3),\\
    &\Omega_{I^-}^{1,1}(s_1s_2;g_1){}^{g_1}\Omega_{I^-}^{1,1}(s_1s_2;g_2)\Omega_{I^-}^{0,2}(g_1,g_2),\\
    &\Omega_{I^-}^{1,1}(s_1s_2s_3;g),
\end{align}
we obtain
\begin{align}
    &d_v\omega^{0,3}=0,\label{1D (0,4)}\\
    &d_h\omega^{0,3}-d_v\omega^{1,2}=0,\label{1D (1,3)}\\
    &d_h\omega^{1,2}+d_v\omega^{2,1}=0,\label{1D (2,2)}\\
    &d_h\omega^{2,1}=0.\label{1D (3,1)}
\end{align}

Therefore, the collection $\omega=\omega^{0,3}+\omega^{1,2}+\omega^{2,1}$ defines a 3-cocycle in the truncated total complex
\begin{equation}
    \tilde{C}^{p,q}:=\begin{cases}
        C^{p,q}&\text{if }q>0,\\
        0&\text{if }q=0.
    \end{cases}
\end{equation}
The redefinitions (\ref{1D (0,3) ambiguity}), (\ref{1D (1,2) ambiguity}), and (\ref{1D (2,1) ambiguity}) are precisely the corresponding 3-coboundary transformations. Hence, the anomaly is characterized by a class
\begin{equation}
    [\omega]\in H^3(\mathrm{Tot}(\tilde{C}^{\bullet,\bullet})).
\end{equation}
As shown in Appendix \ref{Cohomology of Semidirect Products}, this cohomology is identified with the kernel of the restriction map
\begin{equation}
    H^3(\mathrm{Tot}(\tilde{C}^{\bullet,\bullet}))\simeq\ker\left(H^3(B(G\rtimes_\phi G_\mathrm{sp}),U(1))\overset{\pi_*}{\to}H^3(BG_\mathrm{sp},U(1))\right).
\end{equation}
In other words, the anomaly indices are classified by those classes in $H^3(B(G\rtimes_\phi G_\mathrm{sp}),U(1))$ whose restriction to the pure spatial symmetry is trivial. This corresponds to the fact that the spatial symmetry is simply swapping local Hilbert spaces in our setup, and therefore carries no intrinsic anomaly of its own.

\subsection{Example: Lieb-Schultz-Mattis anomaly}\label{Subsec2.2}
A nontrivial Else-Nayak index implies that the corresponding symmetry is not onsite-able and obstructs the existence of a symmetric invertible state. We aim to generalize the statement to the anomaly we obtained in the previous subsection.

For the onsite-ability, there is actually a simple counterexample of an on-site symmetry with nontrivial anomaly. Consider a 1D lattice with a uniform onsite $G$-symmetry
\begin{equation}
    U(g)=\prod_i U_i(g).
\end{equation}
This symmetry has a Lieb-Schultz-Mattis (LSM) anomaly if the representation
\begin{align}
    \begin{split}
        G&\to\mathrm{Aut}(\mathcal{H}_i)\\
        g&\mapsto U_i(g)
    \end{split}
\end{align}
is nontrivially projective, which is described by
\begin{equation}
    U(g_1)U(g_2)=\alpha(g_1,g_2)U(g_1g_2)
\end{equation}
where $[\alpha]\neq0\in H^2(BG,U(1))$. This index can be captured by our framework.

Given a segment $I=[j_1,j_2]$, choose
\begin{equation}
    U_I(g):=\prod_{i\in I} U_i(g),
\end{equation}
then,
\begin{equation}
    \Omega^{0,2}_{\partial I}(g_1,g_2)=\alpha(g_1,g_2)^{j_2-j_1+1}.
\end{equation}
Taking the spatial symmetry to be lattice translation $V(s):=T^s$ for every $s\in\mathbb{Z}$, we have
\begin{equation}
    \Omega^{1,1}_{\partial I}(s;g)=1.
\end{equation}
A convenient choice of endpoint operators is
\begin{align}
    &\Omega^{0,2}_{I^+}(g_1,g_2)=\alpha(g_1,g_2)^{j_2},\quad\Omega^{0,2}_{I^-}(g_1,g_2)=\alpha(g_1,g_2)^{1-j_1},\\
    &\Omega^{1,1}_{I^\pm}(s;g)=1.
\end{align}
The corresponding anomaly indices are
\begin{align}
    &\omega^{0,3}(g_1,g_2,g_3)=\frac{\alpha(g_2,g_3)^{j_2}\alpha(g_1,g_2g_3)^{j_2}}{\alpha(g_1g_2,g_3)^{j_2}\alpha(g_1,g_2)^{j_2}}=1,\\
    &\omega^{1,2}(s;g_1,g_2)=\frac{\alpha(g_1,g_2)^{j_2-s}\cdot1}{\alpha(g_1,g_2)^{j_2}\cdot1\cdot1}=\alpha(g_1,g_2)^{-s},\\
    &\omega^{2,1}(s_1,s_2;g)=\frac{1\cdot1}{1}=1.
\end{align}
Thus, $\omega^{1,2}$ can be nontrivial even though the symmetry is onsite.

Nevertheless, the above index describes the LSM anomaly, which does obstruct the existence of a symmetric invertible state. We are going to prove that a nontrivial index $[\omega]$ obstructs the existence of a symmetric invertible state in general.

\subsection{Dynamical consequence of ‘t Hooft anomaly}
\begin{theorem}\label{Dynamical Consequence 1D}
    Given a finite locality-preserving unitary internal $G$-symmetry $\{U(g)\}$ together with a spatial $G_\mathrm{sp}$-symmetry $\{V(s)\}$ on a 1D lattice, described by a semidirect product $G\rtimes_\phi G_\mathrm{sp}$. If there exists a symmetric invertible state $|\psi\rangle$, then the symmetry has a trivial anomaly index $[\omega]\in\ker\left(H^3(B(G\rtimes_\phi G_\mathrm{sp}),U(1))\overset{\pi_*}{\to}H^3(BG_\mathrm{sp},U(1))\right)$.
\end{theorem}
\begin{proof}
    By Lemma \ref{SRE to Invertible}, it is sufficient to prove that the statement holds for a symmetric state $|\psi\rangle$ that is not only invertible, but also short-range entangled (SRE), i.e.,
\begin{equation}
    U(g)|\psi\rangle=|\psi\rangle,\quad V(s)|\psi\rangle=|\psi\rangle,
\end{equation}
and there is a finite-depth quantum circuit $W$ such that
\begin{equation}
    |\psi\rangle=W|0\rangle,
\end{equation}
where $|0\rangle=\bigotimes_i|0\rangle_i$ is an unentangled tensor product state. Conjugating the symmetry operators by $W$ and define
\begin{equation}
    \tilde{U}(g):=W^\dagger U(g)W.
\end{equation}
Then,
\begin{equation}
    \tilde{U}(g)|0\rangle=|0\rangle.
\end{equation}

By Lemma \ref{1D Truncation}, we can find a truncation $\tilde{U}_{\tilde{I}}(g)$ of $\tilde{U}(g)$ such that $\tilde{U}_{\tilde{I}}(g)|0\rangle=|0\rangle$. We may then define a truncation of $U(g)$ as
\begin{equation}
    U_I(g):=W\tilde{U}_{\tilde{I}}(g)W^\dagger,
\end{equation}
where $I$ is a finite enlargement of $\tilde{I}$ since $W$ is finite-depth. By construction,
\begin{equation}
    U_{I}(g)|\psi\rangle=|\psi\rangle.
\end{equation}
Thus, the endpoints operators act trivially on the symmetric SRE state
\begin{align}
    \begin{split}
        &\Omega^{0,2}_{\partial I}(g_1,g_2)|\psi\rangle=U_I(g_1)U_I(g_2)U_I(g_1g_2)^{-1}|\psi\rangle=|\psi\rangle,\\
        &\Omega^{1,1}_{\partial I}(s;g)|\psi\rangle=V(s)U_{s^{-1}I}(^{s}g)V(s)^{-1}U_I(g)^{-1}|\psi\rangle=|\psi\rangle.
    \end{split}
\end{align}
We can then choose 
\begin{equation}
    \Omega^{0,2}_{I^\pm}(g_1,g_2)|\psi\rangle=|\psi\rangle,\quad\Omega^{1,1}_{I^\pm}(s;g)|\psi\rangle=|\psi\rangle,
\end{equation}
and all $\omega$'s are therefore trivial.
\end{proof}

\begin{lemma}\label{SRE to Invertible}
    Suppose a symmetry with a symmetric SRE state must have a trivial anomaly index. Then, a symmetry with a symmetric invertible state must also have a trivial anomaly index.
\end{lemma}
\begin{proof}
    Given a symmetric invertible state $|\psi\rangle$. By definition, we can find an ancillary state $|\phi\rangle$ such that $|\psi'\rangle:=|\psi\rangle\otimes|\phi\rangle$ is a SRE state. Hence, the symmetry generated by 
    $\{U(g)\otimes\mathbbm{1}\}$ and $\{V(s)\otimes\mathbbm{1}\}$ of $|\psi\rangle\otimes|\phi\rangle$ must have a trivial anomaly index since it is the symmetry of $|\psi'\rangle$. Therefore, the original symmetry generated by 
    $\{U(g)\}$ and $\{V(s)\}$ also has a trivial anomaly index.
\end{proof}

\begin{lemma}\label{1D Truncation}
    Given a 1D FDQC $U$ and an unentangled tensor product state $|0\rangle$ such that $U|0\rangle=|0\rangle$. Given a truncation $U_I$ of $U$ for a sufficiently large segment $I$, there exists a FDQC $\Omega_{\partial I}$ supported near the two endpoints of $I$ such that $\Omega_{\partial I}U_I|0\rangle=|0\rangle$. In other words, we can find a truncation of $U$ that acts trivially on $|0\rangle$.
\end{lemma}
\begin{proof}
    Given a truncation $U_I$ of $U$ for a sufficiently large segment $I$. Since $U|0\rangle=|0\rangle$, its action on $|0\rangle$ can be nontrivial only near the endpoints of $I$. Hence, we may write
    \begin{equation}
    U_I|0\rangle=|\phi\rangle_{\partial I}\otimes\bigotimes_{i\notin\partial I}|0\rangle_i
    \end{equation}
    for $|\phi\rangle_{\partial I}$ defined near the two endpoints of $I$. Let $\Omega_{\partial I}$ be a unitary operator supperted near the endpoints of $I$ such that $\Omega_{\partial I}|\phi\rangle_{\partial I}=|0\rangle_{\partial I}$, we then have $\Omega_{\partial I}U_I|0\rangle=|0\rangle$ as desired.
\end{proof}
\section{Anomalies of 0-Form Symmetries in 2+1D}\label{Sec3}
Consider the same setup as the one in the previous section. The only difference is that $\Lambda$ is now a 2D lattice without boundary.

In this section, we characterize the ‘t Hooft anomaly of this symmetry by generalizing the arguments in \cite{kawagoe2025anomaly}. Detailed calculations are given in Appendix \ref{Conditions}. We further show that a nontrivial ‘t Hooft anomaly obstructs the existence of a symmetric invertible state.

\subsection{GNVW index}
Let $R\subset\Lambda$ be a sufficiently large simply connected region. For each $g\in G$, we choose a truncation $U_R(g)$ of the symmetry operator $U(g)$ such that, for every two regions $R$ and $R'$ whose boundaries have a sufficiently long overlap and the orientations of the two regions match on this overlap, the actions of $U_R(g)$ and $U_{R'}(g)$ are the same near this overlap.

Since $U(g_1)U(g_2)=U(g_1g_2)$, the operator $U_R(g_1)U_R(g_2)U_R(g_1g_2)^{-1}$ is supported near the boundary of $R$. We therefore define
\begin{equation}
    U_R(g_1)U_R(g_2)=\Omega_{\partial R}^{0,2}(g_1,g_2)U_R(g_1g_2),
\end{equation}
where $\Omega_{\partial R}^{0,2}(g_1,g_2)$ is supported near the boundary of $R$.

Similarly, $V(s)U(^sg)V(s)^{-1}=U(g)$ and $V(s)$ maps $x\in\Lambda$ to $sx\in\Lambda$, the operator $V(s)U_{s^{-1}R}(^sg)V(s)^{-1}U_R(g)^{-1}$ is supported near the boundary of $R$. We define
\begin{equation}
    V(s)U_{s^{-1}R}(^sg)=\Omega_{\partial R}^{1,1}(s;g)U_R(g)V(s),
\end{equation}
where $\Omega_{\partial R}^{1,1}(s;g)$ is supported near the boundary of $R$.

We can view $\partial R$ effectively as a 1D lattice, and the boundary operators $\Omega_{\partial R}^{0,2}$ and $\Omega_{\partial R}^{1,1}$ as 1D QCAs. These boundary operators are not finite-depth quantum circuits as 1D QCAs in general. Hence, we define
\begin{align}
    &\nu^{0,2}_{\partial R}(g_1,g_2):=\mathrm{Ind}\left(\Omega_{\partial R}^{0,2}(g_1,g_2)\right),\\
    &\nu^{1,1}_{\partial R}(s;g):=\mathrm{Ind}\left(\Omega_{\partial R}^{1,1}(s;g)\right),
\end{align}
where $\mathrm{Ind}$ is the GNVW index of 1D QCAs~\cite{GNVW2012}.

These indices are independent of the choices of truncations. The justification is as follows. We first consider two regions $R$ and $R'$ whose boundaries have a sufficiently long overlap
and the orientations of the two regions match on this overlap. By construction of the truncated symmetry operators, $U_R(g)U_{R'}(g)^{-1}$ is trivial near the overlap and is supported near $R\Delta R'$, which can be viewed as a 1D QCA with ends. Thus, the GNVW index of this effective 1D QCA $U_R(g)U_{R'}(g)^{-1}$ is trivial, and hence,
\begin{align}
    \begin{split}
        \nu^{0,2}_{\partial R}(g_1,g_2)\nu^{0,2}_{\partial R'}(g_1,g_2)^{-1}&=\mathrm{Ind}\left(\Omega_{\partial R}^{0,2}(g_1,g_2)\Omega_{\partial R'}^{0,2}(g_1,g_2)^{-1}\right)\\
        &=\mathrm{Ind}\left(U_R(g_1)U_R(g_2)U_R(g_1g_2)^{-1}U_{R'}(g_1,g_2)U_{R'}(g_2)^{-1}U_{R'}(g_1)^{-1}\right)=0,
    \end{split}\\
    \begin{split}
        \nu^{1,1}_{\partial R}(s;g)\nu^{1,1}_{\partial R'}(s;g)^{-1}&=\mathrm{Ind}\left(\Omega_{\partial R}^{1,1}(s;g)\Omega_{\partial R'}^{1,1}(s;g)^{-1}\right)\\
        &=\mathrm{Ind}\left(V(s)U_{s^{-1}R}(^sg)V(s)^{-1}U_R(g)^{-1}U_{R'}(g)V(s)U_{s^{-1}R'}(^sg)^{-1}V(s)^{-1}\right)=0.
    \end{split}
\end{align}

The truncation $U_R$ may be modified by operators supported near $\partial R$. Consider the redefinition
\begin{equation}
    U_R(g)\mapsto\Sigma_{\partial R}(g)U_R(g)
\end{equation}
for operators $\Sigma_{\partial R}(g)$ supported near $\partial R$. By construction of the truncated symmetry operators,
\begin{equation}
    \mu^{0,1}(g):=\mathrm{Ind}(\Sigma_{\partial R}(g))
\end{equation}
must be independent of $R$. Under
the redefinition,
\begin{align}
    &\nu^{0,2}(g_1,g_2)\mapsto\nu^{0,2}(g_1,g_2)\mu^{0,1}(g_2)\mu^{0,1}(g_1g_2)^{-1}\mu^{0,1}(g_1),\\
    &\nu^{1,1}(s;g)\mapsto\nu^{1,1}(s;g){}^s\mu^{0,1}({}^sg)\mu^{0,1}(g)^{-1}.
\end{align}
Here, ${}^s\mu^{0,1}({}^sg):=\mu^{0,1}({}^sg)^{r(s)}$. We introduce this $G_{\mathrm{sp}}$ action on $\mathbb{Q}_+$ since the GNVW index will be sent to its multiplicative inverse if it undergoes an orientation flip. Hence,
\begin{equation}
    \mathrm{Ind}({}^s\Sigma_{\partial(s^{-1}R)}({}^sg))={}^s\mathrm{Ind}(\Sigma_{\partial(s^{-1}R)}({}^sg))={}^s\mu^{0,1}({}^sg).
\end{equation}

These relations admit a natural formulation in terms of a double complex. Consider
\begin{equation}
    C^{p,q}:=C^p(BG_{\mathrm{sp}},C^q(BG,\mathbb{Q}_+)),
\end{equation}
with horizontal and vertical differentials $d_h$ and $d_v$, respectively (see Appendix \ref{Cohomology of Semidirect Products}). $\nu^{p,q}$ and $\mu^{p,q}$ naturally define elements in $C^{p,q}$. The above ambiguities then become
\begin{align}
    \nu^{0,2}&\mapsto\nu^{0,2}+d_v\mu^{0,1},\\
    \nu^{1,1}&\mapsto\nu^{1,1}+d_h\mu^{0,1}.
\end{align}

By evaluating
\begin{align}
    &U_R(g_1)U_R(g_2)U_R(g_3),\\
    &V(s)U_{s^{-1}R}(^sg_1)U_{s^{-1}R}(^sg_2),\\
    &V(s_1)V(s_2)U_{(s_1s_2)^{-1}R}({}^{s_1s_2}g_1)
\end{align}
in two different ways, we obtain
\begin{align}
    &\Omega_{\partial R}^{0,2}(g_1,g_2)\Omega_{\partial R}^{0,2}(g_1g_2,g_3)={}^{g_1}\Omega_{\partial R}^{0,2}(g_2,g_3)\Omega_{\partial R}^{0,2}(g_1,g_2g_3),\label{2D 0-form (0,3)}\\
    &\Omega_{\partial R}^{1,1}(s;g_1){}^{g_1}\Omega_{\partial R}^{1,1}(s;g_2)\Omega_{\partial R}^{0,2}(g_1,g_2)={}^s\Omega_{\partial(s^{-1}R)}^{0,2}({}^sg_1,{}^sg_2)\Omega_{\partial R}^{1,1}(s;g_1g_2),\label{2D 0-form (1,2)}\\
    &\Omega_{\partial R}^{1,1}(s_1s_2;g)={}^{s_1}\Omega_{\partial(s_1^{-1}R)}^{1,1}(s_2;{}^{s_1}g)\Omega_{\partial R}^{1,1}(s_1;g).\label{2D 0-form (2,1)}
\end{align}
Here we use the notation
\begin{align}
    &{}^g\mathcal{O}:=U_R(g)\mathcal{O}U_R(g)^{-1},\\
    &{}^s\mathcal{O}:=V(s)\mathcal{O}V(s)^{-1}.
\end{align}
By taking the GNVW index, these equations give
\begin{align}
    &d_v\nu^{0,2}=0,\\
    &d_h\nu^{0,2}-d_v\nu^{1,1}=0,\\
    &d_h\nu^{1,1}=0.
\end{align}

Therefore, the collection $\nu=\nu^{0,2}+\nu^{1,1}$ defines a 2-cocycle in the truncated total complex
\begin{equation}
    \tilde{C}^{p,q}:=\begin{cases}
        C^{p,q}&\text{if }q>0,\\
        0&\text{if }q=0,
    \end{cases}
\end{equation}
with ambiguities precisely described by 2-coboundary transformations. Hence, 
\begin{equation}
    [\nu]\in H^2(\mathrm{Tot}(\tilde{C}^{\bullet,\bullet})).
\end{equation}
As shown in Appendix \ref{Cohomology of Semidirect Products}, this cohomology is identified with the kernel of the restriction map
\begin{equation}
    H^2(\mathrm{Tot}(\tilde{C}^{\bullet,\bullet}))\simeq\ker\left(H^2(B(G\rtimes_\phi G_\mathrm{sp}),\mathbb{Q}_+)\overset{\pi_*}{\to}H^2(BG_\mathrm{sp},\mathbb{Q}_+)\right).
\end{equation}
In other words, the GNVW indices are classified by those classes in $H^2(B(G\rtimes_\phi G_\mathrm{sp}),\mathbb{Q}_+)$ whose restriction to the pure spatial symmetry is trivial. This gives a crystalline equivariant generalization of the index discussed in \cite{kapustin2025anomaly2d,shirley2025QCA, TuLongElse2026, feng2026onsite} that obstructs onsite realization of internal symmetries.

\subsection{‘t Hooft anomaly}
We want to further truncate the boundary operators $\Omega$'s. To do so, we need to define modified boundary operators $\hat{\Omega}$'s with trivial GNVW index canonically. We introduce an ancillary 1D qudit chain $\mathcal{H}_{\partial R}$ living along $\partial R$. For a boundary operator $\Omega$ with GNVW index $\nu$, we aim to define a canonical operator $T_{\nu^{-1}}$ on $\mathcal{H}_{\partial R}$ such that $\mathrm{Ind}(T_{\nu^{-1}})=\nu^{-1}$. Then, $\hat{\Omega}:=\Omega\otimes T_{\nu^{-1}}$ has a trivial GNVW index. We also expect that $T_{\nu}T_{\nu'}=T_{\nu\nu'}$. Consequently, the 2-cocycle conditions of $\nu$'s imply that the equations \eqref{2D 0-form (0,3)}-\eqref{2D 0-form (2,1)} still hold after replacing all $\Omega$'s by $\hat{\Omega}$'s.

The desired canonical operator $T_\nu$ is constructed as follows. Since $\nu\in\mathbb{Q}_+$, we can write down the prime decomposition
\begin{equation}
    \nu=\prod_{k=1}^r{p_k}^{a_k}
\end{equation}
where $p_k$'s are prime numbers and $a_k\in\mathbb{Z}$. Consider an ancillary 1D qudit chain $\mathcal{H}_{\partial R}$ living along $\partial R$ to be a periodic chain with length $L\gg a_k$, and define the ancillary Hilbert space as
\begin{align}
    \begin{split}
        \mathcal{H}_{\partial R}:=\bigotimes_{k=1}^r \mathcal{H}_{p_k},\quad \mathcal{H}_{p_k}:=\bigotimes_{i=1}^L\mathbb{C}^{p_k}.
    \end{split}
\end{align}
The canonical operator $T_\nu$ is then defined as
\begin{equation}
    T_\nu:=\prod_{k=1}^r{t_{p_k}}^{a_k},
\end{equation}
where $t_{p_k}$ denotes the unit lattice translation of $\mathcal{H}_{p_k}$. By construction,
\begin{align}
    &\mathrm{Ind}(T_{\nu})=\prod_{k=1}^r{p_k}^{a_k}=\nu,\\
    &T_{\nu}T_{\nu'}=\prod_{k=1}^r{t_{p_k}}^{a_k}\prod_{l=1}^r{t_{p_l}}^{a'_l}=\prod_{k=1}^r{t_{p_k}}^{a_k+a'_k}=T_{\nu\nu'}.
\end{align}

Let $I\subset\partial R$ be a sufficiently large connected segment and choose a truncation $\Omega_I$ of each extended boundary operators $\hat{\Omega}_{\partial R}$. The equations \eqref{2D 0-form (0,3)}-\eqref{2D 0-form (2,1)} imply that we can define operators $\Gamma_{\partial I}$ localized near the two endpoints of $I$ satisfying
\begin{align}
    &\Omega_{I}^{0,2}(g_1,g_2)\Omega_{I}^{0,2}(g_1g_2,g_3)=\Gamma_{\partial I}^{0,3}(g_1,g_2,g_3){}^{g_1}\Omega_{I}^{0,2}(g_2,g_3)\Omega_{I}^{0,2}(g_1,g_2g_3),\\
    &\Omega_{I}^{1,1}(s;g_1){}^{g_1}\Omega_{I}^{1,1}(s;g_2)\Omega_{I}^{0,2}(g_1,g_2)=\Gamma_{\partial I}^{1,2}(s;g_1,g_2){}^s\Omega_{s^{-1}I}^{0,2}({}^sg_1,{}^sg_2)\Omega_{I}^{1,1}(s;g_1g_2),\\
    &\Omega_{I}^{1,1}(s_1s_2;g)=\Gamma_{\partial I}^{2,1}(s_1,s_2;g){}^{s_1}\Omega_{s_1^{-1}I}^{1,1}(s_2;{}^{s_1}g)\Omega_{I}^{1,1}(s_1;g),
\end{align}
where $\Gamma_{\partial I}=\Gamma_{I^+}\Gamma_{I^-}$ with $\Gamma_{I^+}$ and $\Gamma_{I^-}$ supported near the two endpoints of $I$, respectively. Suppose $\partial R$ and $\partial R'$ have a sufficiently long overlap with the same orientation and $I\subset\partial R$ and $I'\subset\partial R'$ overlap near one of their common endpoint, we choose $\Omega_I$'s consistently such that the actions of $\Omega_I$ and $\Omega_I'$ are the same near their common endpoint and $\Gamma_{I^\pm}=\Gamma_{I'^\pm}$.

Since
\begin{align}
    \begin{split}
        &U_R(g_1)U_R(g_2)U_R(g_3)U_R(g_4)\\
        =&\Omega_{\partial R}^{0,2}(g_1,g_2)\Omega_{\partial R}^{0,2}(g_1g_2,g_3)\Omega_{\partial R}^{0,2}(g_1g_2g_3,g_4)U_R(g_1g_2g_3g_4),
    \end{split}\\
    \begin{split}
        &V(s)U_{s^{-1}R}({}^{s}g_1)U_{s^{-1}R}({}^{s}g_2)U_{s^{-1}R}({}^{s}g_3)\\
        =&\Omega_{\partial R}^{1,1}(s;g_1){}^{g_1}\Omega_{\partial R}^{1,1}(s;g_2)\Omega_{\partial R}^{0,2}(g_1,g_2){}^{g_1g_2}\Omega_{\partial R}^{1,1}(s;g_3)\Omega_{\partial R}^{0,2}(g_1g_2,g_3)U_R(g_1g_2g_3)V(s),
    \end{split}\\
    \begin{split}
        &V(s_1)V(s_2)U_{(s_1s_2)^{-1}R}({}^{s_1s_2}g_1)U_{(s_1s_2)^{-1}R}({}^{s_1s_2}g_2)\\
        =&\Omega_{\partial R}^{1,1}(s_1s_2;g_1){}^{g_1}\Omega_{\partial R}^{1,1}(s_1s_2;g_2)\Omega_{\partial R}^{0,2}(g_1,g_2)U_R(g_1g_2)V(s_1s_2),
    \end{split}\\
    \begin{split}
        &V(s_1)V(s_2)V(s_3)U_{(s_1s_2s_3)^{-1}R}({}^{s_1s_2s_3}g)\\
        =&\Omega_{\partial R}^{1,1}(s_1s_2s_3)(g)U_R(g)V(s_1s_2s_3),
    \end{split}
\end{align}
we can evaluate
\begin{align}
    &\Omega_{I}^{0,2}(g_1,g_2)\Omega_{I}^{0,2}(g_1g_2,g_3)\Omega_{I}^{0,2}(g_1g_2g_3,g_4),\\
    &\Omega_{I}^{1,1}(s;g_1){}^{g_1}\Omega_{I}^{1,1}(s;g_2)\Omega_{I}^{0,2}(g_1,g_2){}^{g_1g_2}\Omega_{I}^{1,1}(s;g_3)\Omega_{I}^{0,2}(g_1g_2,g_3),\\
    &\Omega_{I}^{1,1}(s_1s_2;g_1){}^{g_1}\Omega_{I}^{1,1}(s_1s_2;g_2)\Omega_{I}^{0,2}(g_1,g_2),\\
    &\Omega_{I}^{1,1}(s_1s_2s_3;g)
\end{align}
in two different ways and get
\begin{align}
    \begin{split}
        &\Gamma_{\partial I}^{0,3}(g_1,g_2,g_3){}^{{}^{g_1}(g_2,g_3)}\Gamma_{\partial I}^{0,3}(g_1,g_2g_3,g_4){}^{g_1}\Gamma_{\partial I}^{0,3}(g_2,g_3,g_4)\\
        =&{}^{(g_1,g_2)}\Gamma_{\partial I}^{0,3}(g_1g_2,g_3,g_4)\Delta_{\partial I}^{g_1,g_2,(g_3,g_4)}{}^{{}^{g_1\cdot g_2}(g_3,g_4)}\Gamma_{\partial I}^{0,3}(g_1,g_2,g_3g_4),\label{2D 0-form (0,4)}
    \end{split}\\
    \begin{split}
        &\Gamma_{\partial I}^{1,2}(s;g_1,g_2){}^{{}^{s}({}^{s}g_1,{}^{s}g_2)}\Gamma_{\partial I}^{1,2}(s;g_1g_2,g_3){}^s\Gamma_{\partial(s^{-1}I)}^{0,3}({}^sg_1,{}^sg_2,{}^sg_3)\\
        =&{}^{(s;g_1)}\left({}^{{}^{g_1}(s;g_2)}\Delta_{\partial I}^{g_1,g_2,(s;g_3)}\right){}^{(s;g_1)}\left({}^{{}^{g_1}(s;g_2)}\left({}^{{}^{g_1\cdot g_2}(s;g_3)}\Gamma_{\partial I}^{0,3}(g_1,g_2,g_3)\right)\right){}^{(s;g_1)}\left({}^{g_1}\Gamma_{\partial I}^{1,2}(s;g_2,g_3)\right)\\
        &\quad\Delta_{\partial I}^{g_1,s,({}^sg_2,{}^sg_3)}{}^{{}^{s\cdot {}^sg_1}({}^sg_2,{}^sg_3)}\Gamma_{\partial I}^{1,2}(s;g_1,g_2g_3),\label{2D 0-form (1,3)}
    \end{split}\\
    \begin{split}
        &\Gamma_{\partial I}^{1,2}(s_1s_2;g_1,g_2){}^{{}^{s_1s_2}({}^{s_1s_2}g_1,{}^{s_1s_2}g_2)}\Gamma_{\partial I}^{2,1}(s_1,s_2;g_1g_2)\\
        =&\Gamma_{\partial I}^{2,1}(s_1,s_2;g_1){}^{{}^{s_1}(s_2;{}^{s_1}g_1)}\left({}^{(s_1;g_1)}\left({}^{g_1}\Gamma_{\partial I}^{2,1}(s_1,s_2;g_2)\right)\right){}^{{}^{s_1}(s_2;{}^{s_1}g_1)}\Delta_{\partial I}^{g_1,s_1,(s_2,{}^{s_1}g_2)}\\
        &\quad{}^{{}^{s_1}(s_2;{}^{s_1}g_1)}\left({}^{{}^{s_1\cdot {}^{s_1}g_1}(s_2,{}^{s_1}g_2)}\Gamma_{\partial I}^{1,2}(s_1;g_1,g_2)\right){}^{s_1}\Gamma_{\partial(s_1^{-1}I)}^{1,2}(s_2;{}^{s_1}g_1,{}^{s_1}g_2),\label{2D 0-form (2,2)}
    \end{split}\\
    \begin{split}
        &\Gamma_{\partial I}^{2,1}(s_1s_2,s_3;g){}^{{}^{s_1s_2}(s_3;{}^{s_1s_2}g)}\Gamma_{\partial I}^{2,1}(s_1,s_2;g)\\
        =&\Gamma_{\partial I}^{2,1}(s_1,s_2s_3;g){}^{s_1}\Gamma_{\partial(s_1^{-1}I)}^{2,1}(s_2,s_3;{}^{s_1}g),\label{2D 0-form (3,1)}
    \end{split}
\end{align}
where
\begin{align}
    &{}^{(g_1,g_2)}\mathcal{O}:=\Omega_{I}^{0,2}(g_1,g_2)\mathcal{O}\Omega_{I}^{0,2}(g_1,g_2)^{-1},\\
    &{}^{(s;g)}\mathcal{O}:=\Omega_{I}^{1,1}(s;g)\mathcal{O}\Omega_{I}^{1,1}(s;g)^{-1},\\
    &{}^{{}^{g_1}(g_2,g_3)}\mathcal{O}:={}^{g_1}\Omega_{I}^{0,2}(g_2,g_3)\mathcal{O}{}^{g_1}\Omega_{I}^{0,2}(g_2,g_3)^{-1},\\
    &{}^{{}^{g_1}(s;g_2)}\mathcal{O}:={}^{g_1}\Omega_{I}^{1,1}(s;g_2)\mathcal{O}{}^{g_1}\Omega_{I}^{1,1}(s;g_2)^{-1},\\
    &{}^{{}^{s}({}^{s}g_1,{}^{s}g_2)}\mathcal{O}:={}^{s}\Omega_{s^{-1}I}^{0,2}({}^{s}g_1,{}^{s}g_2)\mathcal{O}{}^{s}\Omega_{s^{-1}I}^{0,2}({}^{s}g_1,{}^{s}g_2)^{-1},\\
    &{}^{{}^{s_1}(s_2;{}^{s_1}g)}\mathcal{O}:={}^{s_1}\Omega_{s_1^{-1}I}^{1,1}(s_2;{}^{s_1}g)\mathcal{O}{}^{s_1}\Omega_{s_1^{-1}I}^{1,1}(s_2;{}^{s_1}g)^{-1},\\
    &{}^{{}^{g_1\cdot g_2}(g_3,g_4)}\mathcal{O}:={}^{g_1}\left({}^{g_2}\Omega_{I}^{0,2}(g_3,g_4)\right)\mathcal{O}{}^{g_1}\left({}^{g_2}\Omega_{I}^{0,2}(g_3,g_4)\right)^{-1}\\
    &{}^{{}^{g_1\cdot g_2}(s;g_3)}\mathcal{O}:={}^{g_1}\left({}^{g_2}\Omega_{I}^{1,1}(s;g_3)\right)\mathcal{O}{}^{g_1}\left({}^{g_2}\Omega_{I}^{1,1}(s;g_3)\right)^{-1}\\
    &{}^{{}^{s\cdot {}^sg_1}({}^sg_2,{}^sg_3)}\mathcal{O}:={}^{s}\left({}^{{}^sg_1}\Omega_{s^{-1}I}^{0,2}({}^sg_2,{}^sg_3)\right)\mathcal{O}{}^{s}\left({}^{{}^sg_1}\Omega_{s^{-1}I}^{0,2}({}^sg_2,{}^sg_3)\right)^{-1}\\
    &{}^{{}^{s_1\cdot {}^{s_1}g_1}(s_2,{}^{s_1}g_2)}\mathcal{O}:={}^{s_1}\left({}^{{}^{s_1}g_1}\Omega_{s_1^{-1}I}^{1,1}(s_2,{}^{s_1}g_2)\right)\mathcal{O}{}^{s_1}\left({}^{{}^{s_1}g_1}\Omega_{s_1^{-1}I}^{1,1}(s_2,{}^{s_1}g_2)\right)^{-1}\\
    &\Delta_{\partial I}^{g_1,g_2,(g_3,g_4)}:={}^{(g_1,g_2)}\left({}^{g_1g_2}\Omega_{I}^{0,2}(g_3,g_4)\right){}^{g_1}\left({}^{g_2}\Omega_{I}^{0,2}(g_3,g_4)\right)^{-1}\\
    &\Delta_{\partial I}^{g_1,g_2,(s;g_3)}:={}^{(g_1,g_2)}\left({}^{g_1g_2}\Omega_{I}^{1,1}(s;g_3)\right){}^{g_1}\left({}^{g_2}\Omega_{I}^{1,1}(s;g_3)\right)^{-1}\\
    &\Delta_{\partial I}^{g_1,s,({}^sg_2,{}^sg_3)}:={}^{(s;g_1)}\left({}^{g_1}\left({}^s\Omega_{s^{-1}I}^{0,2}({}^sg_2,{}^sg_3)\right)\right){}^{s}\left({}^{{}^s g_1}\Omega_{s^{-1}I}^{0,2}({}^sg_2,{}^sg_3)\right)^{-1}\\
    &\Delta_{\partial I}^{g_1,s_1,(s_2,{}^{s_1}g_2)}:={}^{(s_1;g_1)}\left({}^{g_1}\left({}^{s_1}\Omega_{s_1^{-1}I}^{1,1}(s_2,{}^{s_1}g_2)\right)\right){}^{s_1}\left({}^{{}^{s_1} g_1}\Omega_{s_1^{-1}I}^{1,1}(s_2,{}^{s_1}g_2)\right)^{-1}\\
\end{align}

Since both sides of the above equations are all supported near the two endpoints of $I$, define $\omega$'s as some $U(1)$ phases
\begin{align}
    \begin{split}
        \omega^{0,4}(g_1,g_2,g_3,g_4):=&\Gamma_{I^+}^{0,3}(g_1,g_2,g_3){}^{{}^{g_1}(g_2,g_3)}\Gamma_{I^+}^{0,3}(g_1,g_2g_3,g_4){}^{g_1}\Gamma_{I^+}^{0,3}(g_2,g_3,g_4)\\
        &\quad{}^{{}^{g_1\cdot g_2}(g_3,g_4)}\Gamma_{I^+}^{0,3}(g_1,g_2,g_3g_4)^{-1}{\Delta_{I^+}^{g_1,g_2,(g_3,g_4)}}^{-1}{}^{(g_1,g_2)}\Gamma_{I^+}^{0,3}(g_1g_2,g_3,g_4)^{-1}
    \end{split}\\
    \begin{split}
        \omega^{1,3}(s;g_1,g_2,g_3):=&\Gamma_{I^+}^{1,2}(s;g_1,g_2){}^{{}^{s}({}^{s}g_1,{}^{s}g_2)}\Gamma_{I^+}^{1,2}(s;g_1g_2,g_3){}^s\Gamma_{(s^{-1}I)^{r(s)}}^{0,3}({}^sg_1,{}^sg_2,{}^sg_3)\\
        &\quad {}^{{}^{s\cdot {}^sg_1}({}^sg_2,{}^sg_3)}\Gamma_{I^+}^{1,2}(s;g_1,g_2g_3)^{-1}{\Delta_{I^+}^{g_1,s,({}^sg_2,{}^sg_3)}}^{-1}{}^{(s;g_1)}\left({}^{g_1}\Gamma_{I^+}^{1,2}(s;g_2,g_3)\right)^{-1}\\
        &\quad{}^{(s;g_1)}\left({}^{{}^{g_1}(s;g_2)}\left({}^{{}^{g_1\cdot g_2}(s;g_3)}\Gamma_{I^+}^{0,3}(g_1,g_2,g_3)\right)\right)^{-1}{}^{(s;g_1)}\left({}^{{}^{g_1}(s;g_2)}\Delta_{I^+}^{g_1,g_2,(s;g_3)}\right)^{-1}
    \end{split}\\
    \begin{split}
        \omega^{2,2}(s_1,s_2;g_1,g_2):=&\Gamma_{I^+}^{2,1}(s_1,s_2;g_1){}^{{}^{s_1}(s_2;{}^{s_1}g_1)}\left({}^{(s_1;g_1)}\left({}^{g_1}\Gamma_{I^+}^{2,1}(s_1,s_2;g_2)\right)\right){}^{{}^{s_1}(s_2;{}^{s_1}g_1)}\Delta_{I^+}^{g_1,s_1,(s_2,{}^{s_1}g_2)}\\
        &\quad{}^{{}^{s_1}(s_2;{}^{s_1}g_1)}\left({}^{{}^{s_1\cdot {}^{s_1}g_1}(s_2,{}^{s_1}g_2)}\Gamma_{I^+}^{1,2}(s_1;g_1,g_2)\right){}^{s_1}\Gamma_{(s_1^{-1}I)^{r(s_1)}}^{1,2}(s_2;{}^{s_1}g_1,{}^{s_1}g_2)\\
        &\quad{}^{{}^{s_1s_2}({}^{s_1s_2}g_1,{}^{s_1s_2}g_2)}\Gamma_{I^+}^{2,1}(s_1,s_2;g_1g_2)^{-1}\Gamma_{I^+}^{1,2}(s_1s_2;g_1,g_2)^{-1},
    \end{split}\\
    \begin{split}
        \omega^{3,1}(s_1,s_2,s_3;g):=&\Gamma_{I^+}^{2,1}(s_1,s_2s_3;g){}^{s_1}\Gamma_{(s_1^{-1}I)^{r(s_1)}}^{2,1}(s_2,s_3;{}^{s_1}g){}^{{}^{s_1s_2}(s_3;{}^{s_1s_2}g)}\Gamma_{I^+}^{2,1}(s_1,s_2;g)^{-1}\Gamma_{I^+}^{2,1}(s_1s_2,s_3;g)^{-1}.
    \end{split}
\end{align}
These indices fit in the double complex
\begin{equation}
    C^p(BG_{sp}.C^q(BG,U(1))).
\end{equation}

There are four types of ambiguities in this construction. First, we may redefine each onsite Hilbert space by multiplying an on-site unitary map $U$. Same as the arguments for 1+1D, we obtain the same indices after this redefinition.

Second, the truncation $U_R$ can be modified by operators supported near $\partial R$ as $U_R(g)\mapsto\Sigma_{\partial R}(g)U_R(g)$. Due to the definition of the indices, we are in fact dealing with a FDQC $\Sigma_{\partial R}(g)\otimes T_{\mu^{0,1}(g)^{-1}}$. Hence, it can be realized as finitely many times of local modifications, and the resulting $\omega$'s are invariant under changes of truncation.

Third, the truncation $\Omega_I$ can be modified by operators supported near the two endpoints of $I$. Since the modifications at the two endpoints can be made independently, the resulting $\omega$'s are invariant under changes of truncation.

Lastly, we may rescale the endpoint operators by $U(1)$-valued functions:
\begin{align}
    &\Gamma_{I^\pm}^{0,3}(g_1,g_2,g_3)\mapsto\alpha^{0,3}(g_1,g_2,g_3)^{\pm1}\Gamma_{I^\pm}^{0,3}(g_1,g_2,g_3),\\
    &\Gamma_{I^\pm}^{1,2}(s;g_1,g_2)\mapsto\alpha^{1,2}(s;g_1,g_2)^{\pm1}\Gamma_{I^\pm}^{1,2}(s;g_1,g_2),\\
    &\Gamma_{I^\pm}^{2,1}(s_1,s_2;g)\mapsto\alpha^{2,1}(s_1,s_2;g)^{\pm1}\Gamma_{I^\pm}^{2,1}(s_1,s_2;g).
\end{align}
Owing to our consistency convention for the endpoint operators $\Gamma_{I^\pm}$'s, the functions $\alpha$'s are independent of $I$. Under these redefinitions, the anomaly indices transform as follows:
\begin{align}
    &\omega^{0,4}\mapsto\omega^{0,4}+d_v\alpha^{0,3},\\
    &\omega^{1,3}\mapsto\omega^{1,3}+d_h\alpha^{0,3}-d_v\alpha^{1,2},\\
    &\omega^{2,2}\mapsto\omega^{2,2}+d_h\alpha^{1,2}+d_v\alpha^{2,1},\\
    &\omega^{3,1}\mapsto\omega^{3,1}+d_h\alpha^{2,1}.
\end{align}
Moreover, we can obtain the cocycle conditions (see Appendix \ref{Conditions})
\begin{align}
    &d_v\omega^{0,4}=0,\label{2D 0-form (0,5)}\\
    &d_h\omega^{0,4}-d_v\omega^{1,3}=0,\label{2D 0-form (1,4)}\\
    &d_h\omega^{1,3}+d_v\omega^{2,2}=0,\label{2D 0-form (2,3)}\\
    &d_h\omega^{2,2}-d_v\omega^{3,1}=0,\label{2D 0-form (3,2)}\\
    &d_h\omega^{3,1}=0.\label{2D 0-form (4,1)}
\end{align}

Once again, the anomaly index is characterized by a class
\begin{equation}
    [\omega]\in H^4(\mathrm{Tot}(\tilde{C}^{\bullet,\bullet})).
\end{equation}
As shown in Appendix \ref{Cohomology of Semidirect Products}, this cohomology is identified with the kernel of the restriction map
\begin{equation}
    H^4(\mathrm{Tot}(\tilde{C}^{\bullet,\bullet}))\simeq\ker\left(H^4(B(G\rtimes_\phi G_\mathrm{sp}),U(1))\overset{\pi_*}{\to}H^4(BG_\mathrm{sp},U(1))\right).
\end{equation}
In other words, the anomaly indices are classified by those classes in $H^4(B(G\rtimes_\phi G_\mathrm{sp}),U(1))$ whose restriction to the pure spatial symmetry is trivial. This corresponds to the fact that the spatial symmetry is simply swapping local Hilbert spaces in our setup, and therefore carries no intrinsic anomaly of its own.

\subsection{Dynamical consequence of ‘t Hooft anomaly}
We also aim to prove that a nontrivial anomaly index obstructs the existence of a symmetric invertible state for 2D lattice models. The proof directly follows from \cite{kawagoe2025anomaly}.

\begin{theorem}\label{Dynamical Consequence 2D}
    Given a finite locality-preserving unitary internal $G$-symmetry $\{U(g)\}$ together with a spatial $G_\mathrm{sp}$-symmetry $\{V(s)\}$ on a 2D lattice, described by a semidirect product $G\rtimes_\phi G_\mathrm{sp}$. If there exists a symmetric invertible state $|\psi\rangle$, then the symmetry has a trivial anomaly index $[\omega]\in\ker\left(H^4(B(G\rtimes_\phi G_\mathrm{sp}),U(1))\overset{\pi_*}{\to}H^4(BG_\mathrm{sp},U(1))\right)$.
\end{theorem}
\begin{proof}
    By Lemma \ref{SRE to Invertible}, it is sufficient to prove that the statement holds for a symmetric SRE state $|\psi\rangle=W|0\rangle$. Similar to the proof of Theorem \ref{Dynamical Consequence 1D}, define
    \begin{equation}
        \tilde{U}(g):=W^\dagger U(g)W.
    \end{equation}
    Then,
    \begin{equation}
        \tilde{U}(g)|0\rangle=|0\rangle.
    \end{equation}
    
    By Lemma \ref{2D Truncation}, we can find a truncation $\tilde{U}_{\tilde{R}}(g)$ of $\tilde{U}(g)$ such that $\tilde{U}_{\tilde{R}}(g)|0\rangle=|0\rangle$. We may then define a truncation of $U(g)$ as
    \begin{equation}
        U_R(g):=W\tilde{U}_{\tilde{R}}(g)W^\dagger,
    \end{equation}
    where $R$ is a finite enlargement of $\tilde{R}$ since $W$ is finite-depth. By construction, the boundary operators act trivially on the symmetric SRE state
    \begin{align}
        \begin{split}
            &\Omega^{0,2}_{\partial R}(g_1,g_2)|\psi\rangle=U_R(g_1)U_R(g_2)U_R(g_1g_2)^{-1}|\psi\rangle=|\psi\rangle,\\
            &\Omega^{1,1}_{\partial R}(s;g)|\psi\rangle=V(s)U_{s^{-1}R}(^sg)V(s)^{-1}U_R(g)^{-1}|\psi\rangle=|\psi\rangle.
        \end{split}
    \end{align}
    
    Introduce an ancillary 1D qudit chain living along $\partial R$ and define the extended boundary operators $\hat{\Omega}_{\partial R}=\Omega_{\partial R}\otimes T$ with trivial GNVW indices as in the anomaly computation procedure. These extended boundary operators act trivially on the SRE state $|\psi'\rangle:=|\psi\rangle\otimes|0\rangle$. Thus,
    \begin{equation}
        \tilde{\Omega}_{\partial \tilde{R}}:=(W\otimes\mathbbm{1})^\dagger\hat{\Omega}_{\partial R}(W\otimes\mathbbm{1})=W^\dagger\Omega_{\partial R}W\otimes T
    \end{equation}
    act trivially on $|0\rangle\otimes|0\rangle$.
    
    By Lemma \ref{1D Truncation}, we can find a truncation $\tilde{\Omega}_{\tilde{I}}$ of $\tilde{\Omega}_{\partial \tilde{R}}$ that acts trivially on $|0\rangle\otimes|0\rangle$. We may then define a truncation of $\hat{\Omega}_{\partial R}$ as
    \begin{equation}
        \Omega_{I}:=(W\otimes\mathbbm{1})\tilde{\Omega}_{\tilde{I}}(W\otimes\mathbbm{1})^\dagger,
    \end{equation}
    where $I$ is a finite enlargement of $\tilde{I}$ since $W$ is finite-depth. By construction,
    \begin{equation}
        \Omega_{I}|\psi'\rangle=|\psi'\rangle.
    \end{equation}
    Since $\Gamma_{\partial I}$ are defined by products of $\Omega_{I}$, they also act trivially on $|\psi'\rangle$. We can then choose $\Gamma_{I^\pm}|\psi'\rangle=|\psi'\rangle$, and all $\omega$'s are therefore trivial.    
\end{proof}

\begin{lemma}\label{2D Truncation}
    Given a 2D FDQC $U$ and an unentangled tensor product state $|0\rangle$ such that $U|0\rangle=|0\rangle$. Given a truncation $U_R$ of $U$ for a sufficiently large simply connected region $R$, there exists a 1D FDQC $\Omega_{\partial R}$ supported near $\partial R$ such that $\Omega_{\partial R}U_R|0\rangle=|0\rangle$. In other words, we can find a truncation of $U$ that acts trivially on $|0\rangle$.
\end{lemma}
\begin{proof}
    Recall that all 1D invertible states are SRE \cite{kapustin2021}. The strategy is to show that for any truncation $U_R$ of $U$ for a sufficiently large simply connected region $R$, the nontrivial action on $|0\rangle$ near $\partial R$ gives an invertible state supported near $\partial R$.

    Consider two sufficiently large regions $R'\subset R\subset\Lambda$ such that $R\setminus R'$ is a sufficiently wide annulus, and two truncations $U_R$ and $U_{R'}$ of $U$. Since $U|0\rangle=|0\rangle$, we may write
    \begin{align}
        \begin{split}
            &U_R|0\rangle=|\phi\rangle_{\partial R}\otimes\bigotimes_{i\notin\partial R}|0\rangle_i,\\
            &U_{R'}^{-1}U_R|0\rangle=|\phi\rangle_{\partial R}\otimes|\phi'\rangle_{\partial R'}\otimes\bigotimes_{i\notin\partial R\cup\partial R'}|0\rangle_i,
        \end{split}
    \end{align}
    where $|\phi\rangle_{\partial R}$ and $|\phi'\rangle_{\partial R'}$ are defined near $\partial R$ and $\partial R'$, respectively. By definition, 
    \begin{equation}
        U_R^{-1}U_{R'}(|\phi\rangle_{\partial R}\otimes|\phi'\rangle_{\partial R'})=|0\rangle_{\partial R\cup\partial R'}.
    \end{equation}
    Consequently, $|\phi\rangle_{\partial R}$ is invertible, and hence, it is SRE. Let $\Omega_{\partial R}$ be an FDQC supperted near $\partial R$ such that $\Omega_{\partial R}|\phi\rangle_{\partial R}=|0\rangle_{\partial R}$, we then have $\Omega_{\partial R}U_R|0\rangle=|0\rangle$ as desired.
\end{proof}
\section{Anomalies of Symmetries including Generalized Spatial Symmetries}\label{Sec4}
The spatial symmetries considered so far are all swapping local Hilbert spaces. In this section, we consider a more general kind of spatial symmetries.

\subsection{Generalized spatial symmetries and group extensions} Consider a finite internal $G$-symmetry $\{U(g)\}$ and consider a spatial symmetry group $G_{\mathrm{sp}}$. We allow a generalized spatial symmetry action of the form
\begin{equation}
    V(s):=U(s)V_0(s)
\end{equation}
for every $s\in G_{\mathrm{sp}}$, where $U(s)$ is a QCA and $V_0(s)$ is the conventional spatial symmetry action that purely permutes the local Hilbert spaces. Since $V_0(s_1)V_0(s_2)=V_0(s_1s_2)$, closure of the symmetry implies that
\begin{align}
    \begin{split}
        &V(s)U(g)V(s)^{-1}=U(\phi(s)(g)),\\
        &V(s_1)V(s_2)=U(e(s_1,s_2))V(s_1s_2),
    \end{split}
\end{align}
for some $\phi(s)\in\mathrm{Aut}(G)$ and $e(s_1,s_2)\in G$. Assume that $e(s_1,s_2)\in Z(G)$ for every $s_1,s_2\in G_{\mathrm{sp}}$. Then, $e\in Z^2(BG_{\mathrm{sp}},Z(G))$. These relations define a group extension
\begin{equation}
    1\to G\xrightarrow{i}\widetilde{\Gamma}\xrightarrow{p}G_{\mathrm{sp}}\to1\label{extension}
\end{equation}
together with a set-theoretic section $\sigma:G_{\mathrm{sp}}\to\widetilde{\Gamma}$, where $i(g)=U(g)$ and $\sigma(s)=V(s)$. We label an element
$U(g)V(s)\in\tilde G$ by $(g,s)$. The multiplication law is then
\begin{align}
    \begin{split}
        (g_1,s_1)(g_2,s_2)&=U(g_1)V(s_1)U(g_2)V(s_2)=U(g_1)U(\phi(s_1)(g_2))V(s_1)V(s_2)=U(g_1\phi(s_1)(g_2)e(s_1,s_2))V(s_1s_2)\\
        &=(g_1\phi(s_1)(g_2)e(s_1,s_2),s_1s_2).
    \end{split}
\end{align}

With this understading of the group structure of $\widetilde{\Gamma}$, we may perform the original Else-Nayak characterization of the anomaly indices in both 1+1D \cite{ElseNayak2014} and 2+1D \cite{kawagoe2025anomaly}. By taking a truncation of $(g,s)\in\widetilde{\Gamma}$ to be $U_I((g,s))V_0(s)$ ($U_R((g,s))V_0(s)$) where $U_I((g,s))$ ($U_R((g,s))$) is a truncation of $U(g)U(s)$, the argument of \cite{ElseNayak2014,kawagoe2025anomaly} then gives the anomaly classification in terms of $H^{d+2}(B\widetilde{\Gamma},U(1))$ for $d+1$D lattice ($d=1,2$). A more general statement with a rigorous proof for the 1+1D case is provided in \cite{KapustinSopenko2025}.

\subsection{Modified Lieb-Schultz-Mattis theorem}
As in Subsec. \ref{Subsec2.2}, the LSM anomaly of a uniform onsite $G$-symmetry on a $\mathbb{Z}^2$ lattice is determined by the projective representation carried by each unit cell, represented by a class $\alpha\in H^2(BG,U(1))$. In the cohomological description of anomalies developed in Sec. \ref{Sec3}, this LSM anomaly is encoded by
\begin{equation}
    [\omega^{2,2}]=u\otimes \alpha\in H^2(B\mathbb{Z}^2,H^2(BG,U(1))),
\end{equation}
where $u=u_x\smile u_y$ is a generator of $H^2(B\mathbb{Z}^2,\mathbb{Z})\simeq\mathbb{Z}$, represented by the normalized 2-cocycle
\begin{equation}
    u((x_1,y_1),(x_2,y_2))=x_1y_2.
\end{equation}

We now consider a modification of the LSM theorem arising when translations are realized as generalized spatial symmetries. The result below is an instance of an SPT--LSM theorem, where a nontrivial projective representation per unit cell need not forbid symmetric invertible phases, but can instead require any such phase to have a nontrivial SPT response.

In the cohomological description, this reflects the distinction between an LSM-type class on the second page of the Lyndon--Hochschild--Serre (LHS) spectral sequence described later,
\begin{equation}
E_2^{p,q}=H^p(BG_{\mathrm{sp}},H^q(BG,U(1))),
\end{equation}
and the anomaly of the full symmetry action. A nonzero class on this page can be trivialized by an incoming differential and therefore does not by itself imply a nontrivial total anomaly. Below, we demonstrate this cancellation mechanism explicitly using the microscopic anomaly indices.

In particular, consider an internal symmetry ${U(g)}_{g\in G}$ together with "magnetic" translations $T_x,T_y$, which combine QCAs with translations of the local Hilbert spaces and satisfy
\begin{equation}
T_xT_yT_x^{-1}T_y^{-1}=U(a)
\label{magnetic translations}
\end{equation}
for some $a\in G$. We assume that both translations commute with the internal symmetry, which implies that $a\in Z(G)$.

Under these assumptions, the modified LSM theorem in \cite{Yang2018} can be stated as follows.
\begin{theorem}
    Given a uniform onsite $G$-symmetry $\{U(g)\}$ together with a magnetic translation symmetry satisfying \eqref{magnetic translations} on a $\mathbb{Z}^2$ lattice. Let $\alpha\in H^2(BG,U(1))$ denote its LSM anomaly. If there exists a symmetric invertible state, then there exists a 3-cocycle $\lambda\in Z^3(BG,U(1))$ such that
    \begin{equation}
        [\delta^3_{a}\lambda]=\alpha\quad\text{in } H^2(BG,U(1)),
    \end{equation}
    where $\delta^k_g:C^{k}(BG,U(1))\to C^{k-1}(BG,U(1))$ is defined by
    \begin{equation}
        \delta^k_gf(g_1,\dots g_{k-1}):=\sum_{i=0}^{k-1}(-1)^if(g_1,\dots,g_i,g,g_{i+1},\dots,g_{k-1}).\label{slant products}
    \end{equation}
    That is, the LSM anomaly $H^2(G, U(1))/\text{Im}(\delta_g^3)$ defines an obstruction to symmetric invertible states.
\end{theorem}

\begin{proof}
To prove the theorem, we identify the LSM anomaly with a class on the second page of the LHS spectral sequence and determine the condition for this class to vanish at the limiting page.

\paragraph{The LHS spectral sequence.} For the extension \eqref{extension}, there is a convergence of the LHS spectral sequence 
\begin{equation}
    E_2^{p,q}=H^p(BG_{\mathrm{sp}},H^q(BG,U(1)))\implies H^{p+q}(B\widetilde{\Gamma},U(1)).
\end{equation}
The LSM anomaly lies in $E^{2,2}_2=H^{2}(BG_{\mathrm{sp}},H^2(BG,U(1)))$. Having a symmetric invertible state requires the corresponding anomaly class to vanish in the limiting page. Since $G_{\mathrm{sp}}=\mathbb{Z}^2$ has cohomological dimension 2, all columns with $p>2$ vanish. Consequently, the differentials
\begin{equation}
    d_r^{p,q}:E_r^{p,q}\to E_r^{p+r,q-r+1}
\end{equation}
vanish for $r>2$. Thus, the only possible differential that can remove the LSM anomaly is 
\begin{equation}
    d_2^{0,3}:H^3(BG,U(1))^{G_{\mathrm{sp}}}\to H^2(BG_{\mathrm{sp}},H^2(BG,U(1))),
\end{equation}
and there exists a symmetric invertible state only if the LSM anomaly lies in the image of $d_2^{0,3}$.

\paragraph{The differential \(d_2^{0,k}\).} We now construct this differential
\begin{equation}
    d_2^{0,k}:H^k(BG,U(1))^{G_{\mathrm{sp}}}\to H^2(BG_{\mathrm{sp}},H^{k-1}(BG,U(1)))
\end{equation}
explicitly in terms of homogeneous group cochains. See Appendix \ref{Spectral Sequences} for more details on the differentials in spectral sequences. Let $\hat{\lambda}_0\in C^k(BG,U(1))$ be a homogeneous $k$-cocycle representing a class $[\hat{\lambda}_0]\in H^k(BG,U(1))^{G_{\mathrm{sp}}}$. Homogeneous $k$-cochains are functions $\hat{f}:G^{k+1}\to U(1)$ satisfying
\begin{equation}
    \hat{f}(gg_0,\dots,gg_k)={}^{i(g)}\hat{f}(g_0,\dots,g_k)
\end{equation}
for all $g\in G$, where $c\mapsto{}^{\tilde{g}}c$ is the $\widetilde{\Gamma}$-action on $U(1)$. The corresponding inhomogeneous $k$-cochain representation $f\in G^k\to U(1)$ is obtained by $f(g_1,\dots,g_k)=\hat{f}(1,g_1,g_1g_2,\dots,g_1g_2\dots g_k)$.

By definition, 
\begin{align}
    \begin{split}
        &d_v\hat{\lambda}_0=0,\\
        &[d_h\hat{\lambda}_0]=0,
    \end{split}
\end{align}
where
\begin{align}
    \begin{split}
        &d_v\hat{f}(g_0,\dots,g_{k+1}):=\sum_{i=0}^{k+1}(-1)^i\hat{f}(g_0,\dots,\hat{g_i},\dots,g_{k+1}),\\
        &d_h\hat{f}(s;g_0,\dots,g_k):=T_s\hat{f}(g_0,\dots,g_k)-\hat{f}(g_0,\dots,g_k),
    \end{split}
\end{align}
and $T_s:H^k(BG,U(1))\to H^k(BG,U(1))$ is defined by
\begin{equation}
    T_s\hat{f}(g_0,\dots,g_k):={}^{\sigma(s)}\hat{f}(\phi(s)^{-1}g_0,\dots,\phi(s)^{-1}g_k).
\end{equation}
Although $T_s$ need not define a group action on cochains, it induces the $G_{\mathrm{sp}}$-action on the cohomology classes. Since $[(T_s-\mathrm{id})\hat{\lambda}]=0$, we may choose $\hat{\lambda}_{1,s}\in C^{k-1}(BG,U(1))$ such that
\begin{equation}
    d_v\hat{\lambda}_{1,s}=(T_s-\mathrm{id})\hat{\lambda}_0
\end{equation}
for every $s\in G_{\mathrm{sp}}$.

The failure of the maps $T_s$ to form a group action at the cochain level is controlled by the extension 2-cocycle $e$. In fact, 
\begin{align}
    \begin{split}
        T_{s_1}T_{s_2}\hat{f}(g_0,\dots,g_k)&={}^{\sigma(s_1)\sigma(s_2)}\hat{f}(\phi(s_2)^{-1}\phi(s_1)^{-1}g_0,\dots,\phi(s_2)^{-1}\phi(s_1)^{-1}g_k)\\
        &={}^{i(e(s_1,s_2))}T_{s_1s_2}\hat{f}(e(s_1,s_2)^{-1}g_0e(s_1,s_2),\dots,e(s_1,s_2)^{-1}g_ke(s_1,s_2))\\
        &=T_{e(s_1,s_2)}T_{s_1s_2}\hat{f}(g_0,\dots,g_k),\label{T composition}
    \end{split}
\end{align}
where, for $g\in G$, 
\begin{equation}
    T_g\hat{f}(g_0,\dots,g_k):={}^{i(g)}\hat{f}(g^{-1}g_0g,\dots,g^{-1}g_kg)=\hat{f}(g_0g,\dots,g_kg).
\end{equation}

When computing the horizontal differential of $\hat{\lambda}_1$, the failure of $T_s$ to define a cochain-level action must be compensated by the cochain homotopy $T_{s_1}T_{s_2}-T_{s_1s_2}$. In particular, we need to replace $d_v\hat{\lambda}_{1,s_1s_2}=(T_{s_1s_2}-\mathrm{id})\hat{\lambda}_0$ by $(T_{s_1}T_{s_2}-\mathrm{id})\hat{\lambda}_0=d_v\hat{\lambda}_{1,s_1s_2}+(T_{s_1}T_{s_2}-T_{s_1s_2})\hat{\lambda}_0$. The cochain homotopy between $T_{e(s_1,s_2)}$ and the identity is exactly given by $\delta_{e(s_1,s_2)}$ defined in \eqref{slant products}. More precisely, in homogeneous cochain notation, $\delta_{g}^k:C^k(BG,U(1))\to C^{k-1}(BG,U(1))$ has the form
\begin{align}
    \delta_g^k\hat{f}(g_0,\dots,g_{k-1})=&{}^{g_0}\delta_g^k\hat{f}(1,g_0^{-1}g_1,\dots,g_0^{-1}g_{k-1})={}^{g_0}\delta_g^kf(g_0^{-1}g_1,\dots,g_{k-2}^{-1}g_{k-1})\\
    =&\sum_{i=0}^{k-1}(-1)^i{}^{g_0}f(g_0^{-1}g_1,\dots,g_{i-1}^{-1}g_i,g,g_i^{-1}g_{i+1},\dots,g_{k-2}^{-1}g_{k-1})\\
    =&\sum_{i=0}^{k-1}(-1)^i\hat{f}(g_0,\dots,g_i,g_ig,\dots g_{k-1}g)
\end{align}
for every $g\in Z(G)$. By evaluating $d_v\delta_g^{k}+\delta_g^{k+1}d_v$, we obtain
\begin{align}
    \begin{split}
        &(d_v\delta_g^{k}+\delta_g^{k+1}d_v)\hat{f}(g_0,\dots,g_k)\\
        =&\sum_{i=0}^k(-1)^i\delta_g^{k}\hat{f}(\dots,\hat{g_i},\dots)+\sum_{j=0}^k(-1)^jd_v\hat{f}(\dots,g_j,g_jg,\dots)\\
        =&\sum_{i=0}^k\sum_{j=0}^{i-1}(-1)^{i+j}\hat{f}(\dots,g_j,g_jg,\dots,\hat{g_ig},\dots)+\sum_{i=0}^k\sum_{j=i}^{k-1}(-1)^{i+j}\hat{f}(\dots,\hat{g_i},\dots,g_{j+1},g_{j+1}g,\dots)\\
        &\quad+\sum_{j=0}^k\sum_{i=0}^j(-1)^{i+j}\hat{f}(\dots,\hat{g_i},\dots,g_j,g_jg,\dots)+\sum_{j=0}^k\sum_{i=j+1}^{k+1}(-1)^{i+j}\hat{f}(\dots,g_j,g_jg,\dots,\hat{g_{i-1}g},\dots)\\
        =&\sum_{i=0}^k\sum_{j=0}^{i-1}(-1)^{i+j}\hat{f}(\dots,g_j,g_jg,\dots,\hat{g_ig},\dots)+\sum_{i=0}^k\sum_{j=i+1}^{k}(-1)^{i+j-1}\hat{f}(\dots,\hat{g_i},\dots,g_{j},g_{j}g,\dots)\\
        &\quad+\sum_{j=0}^k\sum_{i=0}^j(-1)^{i+j}\hat{f}(\dots,\hat{g_i},\dots,g_j,g_jg,\dots)+\sum_{j=0}^k\sum_{i=j}^{k}(-1)^{i+j+1}\hat{f}(\dots,g_j,g_jg,\dots,\hat{g_{i}g},\dots)\\
        =&\sum_{i=0}^k\left[\hat{f}(\dots,g_{i-1},g_ig,\dots)-\hat{f}(\dots,g_i,g_{i+1}g,\dots)\right]\\
        =&(T_g-\mathrm{id})\hat{f}(g_0,\dots,g_k).\label{Cochain homotopy}
    \end{split}
\end{align}
Combining \eqref{T composition} and \eqref{Cochain homotopy}, we obtain
\begin{align}
    \begin{split}
        (T_{s_1}T_{s_2}-T_{s_1s_2})\hat{\lambda}_0&=(T_{e(s_1,s_2)}-\mathrm{id})T_{s_1s_2}\hat{\lambda}_0=(d_v\delta_{e(s_1,s_2)}^{k}+\delta_{e(s_1,s_2)}^{k+1}d_v)T_{s_1s_2}\hat{\lambda}_0\\
        &=d_v\delta_{e(s_1,s_2)}^{k}T_{s_1s_2}\hat{\lambda}_0,
    \end{split}
\end{align}
where we used $d_v\hat{\lambda}_0=0$ and the fact that $T_s$ commutes with $d_v$. Thus, we need to replace $\hat{\lambda}_{1,s_1s_2}$ by $\hat{\lambda}_{1,s_1s_2}+\delta_{e(s_1,s_2)}^{k}T_{s_1s_2}\hat{\lambda}_0$ when performing the horizontal differential of $\hat{\lambda}_1$. A representative of $d_2^{0,k}[\hat{\lambda}_0]$ is then given by
\begin{equation}
    d_2^{0,k}[\hat{\lambda}_0](s_1,s_2)=d_h\hat{\lambda}_1(s_1,s_2)=T_{s_1}\hat{\lambda}_{1,s_2}-(\hat{\lambda}_{1,s_1s_2}+\delta_{e(s_1,s_2)}^{k}T_{s_1s_2}\hat{\lambda}_0)+\hat{\lambda}_{1,s_1}.\label{d_2}
\end{equation}

\paragraph{Specialization to magnetic translations.} 
For the magnetic-translation symmetry, the $\mathbb{Z}^2$ translations commute with the internal symmetry and act trivially on the $U(1)$ coefficients since it is orientation-preserving. The induced maps $T_s$ then act trivially on homogeneous $G$-cochains. We may therefore choose $\hat{\lambda}_{1,s}=0$ for every $s\in\mathbb{Z}^2$. With the convention in \eqref{magnetic translations}, 
\begin{equation}
    e((x_1,y_1),(x_2,y_2))=a^{y_1x_2}.
\end{equation}
The representative of $d_2^{0,k}[\hat{\lambda}_0]$ in \eqref{d_2} consequently reduces to
\begin{equation}
    d_2^{0,k}[\hat{\lambda}_0]((x_1,y_1),(x_2,y_2))=-\delta_{a^{y_1x_2}}^{k}\hat{\lambda}_0.
\end{equation}

The $\mathbb{Z}^2$ translations act trivially on $G$-cohomologies also implies that
\begin{align}
    \begin{split}
        E_2^{2,k-1}=H^2(B\mathbb{Z}^2,H^{k-1}(BG,U(1)))&\overset{\sim}{\to}H^{k-1}(BG,U(1)).\\
        u\otimes \omega&\mapsto \omega
    \end{split}
\end{align}
Therefore,
\begin{equation}
    d_2^{0,k}[\hat{\lambda}_0]=(u_y\smile u_x)\otimes[-\delta_a^k\hat{\lambda}_0]=u\otimes[\delta_a^k\hat{\lambda}_0].
\end{equation}
Taking $k=3$, the LSM anomaly $u\otimes\alpha\in E_2^{2,2}$ is in the image of $d_2^{0,3}$ if and only if there exists a 3-cocycle $\lambda\in Z^3(BG,U(1))$ such that
\begin{equation}
    \alpha=[\delta_a^3\lambda]\quad\text{in }H^2(BG,U(1)),
\end{equation}
which completes the proof.
\end{proof}
\section{Discussion}
We close our paper with possible future works. 
An important direction is to extend the anomaly indices to fermionic systems and establish the crystalline equivalence principle directly for microscopic fermionic anomaly indices. The fermionic crystalline equivalence principle has been formulated and investigated in classifications of invertible phases~\cite{ChengWang2022,Debray2021,ManjunathCalveraBarkeshli2023}. Its microscopic characterization is subtler because spatial rotations can act on fermions through a double cover, with a $2\pi$ spinor rotation equal to fermion parity $(-1)^F$. 
Accordingly, representing an IR action of crystalline symmetry by an internal symmetry generally changes the group structure since it shifts the central extension by fermion parity.  It would be valuable to derive how this change affects the anomaly classification directly from the lattice symmetry operators,. Recent constructions of fermionic lattice anomaly indices for internal symmetries would provide a starting point~\cite{ChavdaKobayashi2026}. Incorporating spatial symmetries into these constructions could clarify how the fermionic anomaly layers enter the microscopic equivalence principle.

Another direction concerns the responses and anomalies associated with crystalline defects. Disclinations can bind quantum numbers such as fractional charge, which are constrained by the bulk topological state and symmetry~\cite{ManjunathBarkeshli2021,ZhangManjunathNambiarBarkeshli2022,ManjunathCalveraBarkeshli2023,Kobayashi2025FCI}. Recent work has also related crystalline defects to continuum symmetry defects and associated magnetic fluxes at Dirac criticality~\cite{BarkeshliFechisinKomargodskiZhong2026}. It would be interesting to develop an operator formulation of these phenomena within the present framework. One possible approach is to construct symmetry operators with disclinations by truncating and blending the symmetry operators along the branch cut of the lattice, and then extract the symmetry algebra in the presence of the disclination. This could relate microscopic crystalline anomaly indices to projective symmetry actions and quantum numbers bound at disclinations. A complete prediction of defect responses should not only incorporate the symmetry alone, but the topological data of the state: an anomaly index of the symmetry operators alone would not determine the quantum numbers bound to a defect. Establishing which quantities are fixed by the anomaly, and which depend on the choice of the state, would provide a useful microscopic characterization of crystalline defect responses.

Finally, analogues of SPT--LSM constraints can arise even for purely internal symmetries. A recent example is an exact, non-onsite, exponentially quasi-local $\mathbb Z_4^F$ symmetry of a $p+ip$ superconductor, with its generator satisfying $U^2=(-1)^F$~\cite{KobayashiInamuraShiozaki2026}. Its nontrivial lattice anomaly index excludes symmetric invertible phases with integer chiral central charge, while allowing a shifted family with $c_-\in\mathbb Z+\tfrac12$, including the $p+ip$ state with $c_-=\tfrac12$. This provides an internal-symmetry analogue of the SPT--LSM mechanism: the microscopic symmetry action enforces nontrivial invertible state while remaining compatible with a symmetric invertible phase. It would be interesting to determine which lattice anomaly indices enforce particular chiral or topological responses, classify the compatible invertible phases, and investigate how these constraints combine with crystalline symmetry and its defects.

\section*{Acknowledgments}
We thank Maissam Barkeshli, Meng Cheng, Nathan Seiberg, Sahand Seifnashri and Ken Shiozaki for discussions. We thank Sal Pace for comments on the draft.
R.K. is supported by JST PRESTO Grant Number JPMJPR26A8, Japan, JSPS KAKENHI Grant-in-Aid No.26K24517, and the Department of Applied Physics, the University of Tokyo. 
C-Y.Y. is supported by the GSGC Program, University of Tokyo.
\appendix
\section{Proof of the cocycle conditions}\label{Conditions}
In this appendix, we provide detailed derivations of the cocycle conditions of the anomaly indices. For clarity of the derivation, we neglect the subscript that indicates the support of each boundary and endpoint operator, and denote
\begin{align}
    \begin{split}
        &\Omega_{g_1,g_2}:=\Omega^{0,2}(g_1,g_2),\\
        &\Omega_{s;g}:=\Omega^{1,1}(s;g),\\
        &\Gamma_{g_1,g_2,g_3}:=\Gamma^{0,3}(g_1,g_2,g_3),\\
        &\Gamma_{s;g_1,g_2}:=\Gamma^{1,2}(s;g_1,g_2),\\
        &\Gamma_{s_1,s_2;g}:=\Gamma^{2,1}(s_1,s_2;g).
    \end{split}
\end{align}
 
\subsection{1+1D}
\subsubsection{Definition of the indices}
\begin{align}
    U_I(g_1)U_I(g_2)U_I(g_3)&=\Omega_{g_1,g_2}U_I(g_1g_2)U_I(g_3)=\Omega_{g_1,g_2}\Omega_{g_1g_2,g_3},U_I(g_1g_2g_3)\notag\\
    &={}^{g_1}\Omega_{g_2,g_3}U_I(g_1)U_I(g_2g_3)={}^{g_1}\Omega_{g_2,g_3}\Omega_{g_1,g_2g_3},U_I(g_1g_2g_3),\\
    V(s)U_{s^{-1}I}({}^sg_1)U_{s^{-1}I}({}^sg_2)&=\Omega_{s;g_1}U_I(g_1)V(s)U_{s^{-1}I}({}^sg_2)=\Omega_{s;g_1}{}^{g_1}\Omega_{s;g_2}U_I(g_1)U_I(g_2)V(s)\notag\\
    &=\Omega_{s;g_1}{}^{g_1}\Omega_{s;g_2}\Omega_{g_1,g_2}U_I(g_1g_2)V(s)\notag\\
    &={}^s\Omega_{{}^sg_1,{}^sg_2}V(s)U_{s^{-1}I}({}^s(g_1g_2))={}^s\Omega_{{}^sg_1,{}^sg_2}\Omega_{s;g_1g_2}U_I(g_1g_2)V(s),\\
    V(s_1)V(s_2)U_{(s_1s_2)^{-1}I}({}^{s_1s_2}g)&=\Omega_{s_1s_2;g}U_I(g)V(s_1s_2)\notag\\
    &={}^{s_1}\Omega_{s_2;{}^{s_1}g}V(s_1)U_{s_1^{-1}I}({}^{s_1}g)V(s_2)={}^{s_1}\Omega_{s_2;{}^{s_1}g}\Omega_{s_1;g}U_I(g)V(s_1s_2).
\end{align}
Comparing the two evaluations gives \eqref{1D (0,3)}-\eqref{1D (2,1)}. Keep only the operators supported near the left endpoint $I^-$. The definition of $\omega$'s are given by
\begin{align}
    \begin{split}
        \Omega_{g_1,g_2}\Omega_{g_1g_2,g_3}&:=\omega^{0,3}(g_1,g_2,g_3){}^{g_1}\Omega_{g_2,g_3}\Omega_{g_1,g_2g_3},\\
        \Omega_{s;g_1}{}^{g_1}\Omega_{s;g_2}\Omega_{g_1,g_2}&:=\omega^{1,2}(s;g_1,g_2){}^s\Omega_{{}^sg_1,{}^sg_2}\Omega_{s;g_1g_2},\\
        \Omega_{s_1s_2;g}&:=\omega^{2,1}(s_1,s_2;g){}^{s_1}\Omega_{s_2;{}^{s_1}g}\Omega_{s_1;g}.
    \end{split}
\end{align}
 
\subsubsection{The cocycle conditions}
\paragraph{(0,4)} Take $X_1=\Omega_{g_1,g_2}\Omega_{g_1g_2,g_3}\Omega_{g_1g_2g_3,g_4}$ and $X_2:={}^{g_1}({}^{g_2}\Omega_{g_3,g_4}){}^{g_1}\Omega_{g_2,g_3g_4}\Omega_{g_1,g_2g_3g_4}$.
\begin{align}
    \begin{split}
    X_1&=\omega^{0,3}(g_1,g_2,g_3){}^{g_1}\Omega_{g_2,g_3}\Omega_{g_1,g_2g_3}\Omega_{g_1g_2g_3,g_4}\\
    &=\omega^{0,3}(g_1,g_2,g_3)\omega^{0,3}(g_1,g_2g_3,g_4){}^{g_1}\left(\Omega_{g_2,g_3}\Omega_{g_2g_3,g_4}\right)\Omega_{g_1,g_2g_3g_4}\\
    &=\omega^{0,3}(g_1,g_2,g_3)\omega^{0,3}(g_1,g_2g_3,g_4)\omega^{0,3}(g_2,g_3,g_4)X_2,\\
    X_1&=\omega^{0,3}(g_1g_2,g_3,g_4)\Omega_{g_1,g_2}{}^{g_1g_2}\Omega_{g_3,g_4}\Omega_{g_1g_2,g_3g_4}\\
    &=\omega^{0,3}(g_1g_2,g_3,g_4){}^{g_1}({}^{g_2}\Omega_{g_3,g_4})\Omega_{g_1,g_2}\Omega_{g_1g_2,g_3g_4}\\
    &=\omega^{0,3}(g_1g_2,g_3,g_4)\omega^{0,3}(g_1,g_2,g_3g_4)X_2.
    \end{split}
\end{align}
This gives $d_v\omega^{0,3}=0$.
 
\paragraph{(1,3)} Take $X_1=\Omega_{s;g_1}{}^{g_1}\Omega_{s;g_2}\Omega_{g_1,g_2}{}^{g_1g_2}\Omega_{s;g_3}\Omega_{g_1g_2,g_3}$ and $X_2:={}^s({}^{{}^sg_1}\Omega_{{}^sg_2,{}^sg_3}){}^s\Omega_{{}^sg_1,{}^s(g_2g_3)}\Omega_{s;g_1g_2g_3}$.
\begin{align}
    \begin{split}
    X_1&=\omega^{1,2}(s;g_1,g_2){}^s\Omega_{{}^sg_1,{}^sg_2}\Omega_{s;g_1g_2}{}^{g_1g_2}\Omega_{s;g_3}\Omega_{g_1g_2,g_3}\\
    &=\omega^{1,2}(s;g_1,g_2)\omega^{1,2}(s;g_1g_2,g_3){}^s\left(\Omega_{{}^sg_1,{}^sg_2}\Omega_{{}^s(g_1g_2),{}^sg_3}\right)\Omega_{s;g_1g_2g_3}\\
    &=\omega^{1,2}(s;g_1,g_2)\omega^{1,2}(s;g_1g_2,g_3){}^s\omega^{0,3}({}^sg_1,{}^sg_2,{}^sg_3)X_2,\\
    X_1&=\Omega_{s;g_1}{}^{g_1}\Omega_{s;g_2}{}^{g_1}({}^{g_2}\Omega_{s;g_3})\Omega_{g_1,g_2}\Omega_{g_1g_2,g_3}\\
    &=\omega^{0,3}(g_1,g_2,g_3)\Omega_{s;g_1}{}^{g_1}\left(\Omega_{s;g_2}{}^{g_2}\Omega_{s;g_3}\Omega_{g_2,g_3}\right)\Omega_{g_1,g_2g_3}\\
    &=\omega^{0,3}(g_1,g_2,g_3)\omega^{1,2}(s;g_2,g_3)\Omega_{s;g_1}{}^{g_1}({}^s\Omega_{{}^sg_2,{}^sg_3}){}^{g_1}\Omega_{s;g_2g_3}\Omega_{g_1,g_2g_3}\\
    &=\omega^{0,3}(g_1,g_2,g_3)\omega^{1,2}(s;g_2,g_3){}^s({}^{{}^sg_1}\Omega_{{}^sg_2,{}^sg_3})\Omega_{s;g_1}{}^{g_1}\Omega_{s;g_2g_3}\Omega_{g_1,g_2g_3}\\
    &=\omega^{0,3}(g_1,g_2,g_3)\omega^{1,2}(s;g_2,g_3)\omega^{1,2}(s;g_1,g_2g_3)X_2.
    \end{split}
\end{align}
This gives $d_h\omega^{0,3}-d_v\omega^{1,2}=0$.
 
\paragraph{(2,2)} Take $X_1=\Omega_{s_1s_2;g_1}{}^{g_1}\Omega_{s_1s_2;g_2}\Omega_{g_1,g_2}$ and $X_2:={}^{s_1s_2}\Omega_{{}^{s_1s_2}g_1,{}^{s_1s_2}g_2}{}^{s_1}\Omega_{s_2;{}^{s_1}(g_1g_2)}\Omega_{s_1;g_1g_2}$.
\begin{align}
    \begin{split}
    X_1&=\omega^{1,2}(s_1s_2;g_1,g_2){}^{s_1s_2}\Omega_{{}^{s_1s_2}g_1,{}^{s_1s_2}g_2}\Omega_{s_1s_2;g_1g_2}\\
    &=\omega^{1,2}(s_1s_2;g_1,g_2)\omega^{2,1}(s_1,s_2;g_1g_2)X_2,\\
    X_1&=\omega^{2,1}(s_1,s_2;g_1)\omega^{2,1}(s_1,s_2;g_2){}^{s_1}\Omega_{s_2;{}^{s_1}g_1}\Omega_{s_1;g_1}{}^{g_1}({}^{s_1}\Omega_{s_2;{}^{s_1}g_2}){}^{g_1}\Omega_{s_1;g_2}\Omega_{g_1,g_2}\\
    &=\omega^{2,1}(s_1,s_2;g_1)\omega^{2,1}(s_1,s_2;g_2){}^{s_1}\Omega_{s_2;{}^{s_1}g_1}{}^{s_1}({}^{{}^{s_1}g_1}\Omega_{s_2;{}^{s_1}g_2})\Omega_{s_1;g_1}{}^{g_1}\Omega_{s_1;g_2}\Omega_{g_1,g_2}\\
    &=\omega^{2,1}(s_1,s_2;g_1)\omega^{2,1}(s_1,s_2;g_2)\omega^{1,2}(s_1;g_1,g_2){}^{s_1}\left(\Omega_{s_2;{}^{s_1}g_1}{}^{{}^{s_1}g_1}\Omega_{s_2;{}^{s_1}g_2}\Omega_{{}^{s_1}g_1,{}^{s_1}g_2}\right)\Omega_{s_1;g_1g_2}\\
    &=\omega^{2,1}(s_1,s_2;g_1)\omega^{2,1}(s_1,s_2;g_2)\omega^{1,2}(s_1;g_1,g_2){}^{s_1}\omega^{1,2}(s_2;{}^{s_1}g_1,{}^{s_1}g_2)X_2.
    \end{split}
\end{align}
This gives $d_h\omega^{1,2}+d_v\omega^{2,1}=0$.
 
\paragraph{(3,1)} Take $X_1=\Omega_{s_1s_2s_3;g}$ and $X_2:={}^{s_1s_2}\Omega_{s_3;{}^{s_1s_2}g}{}^{s_1}\Omega_{s_2;{}^{s_1}g}\Omega_{s_1;g}$.
\begin{align}
    \begin{split}
    X_1&=\omega^{2,1}(s_1s_2,s_3;g){}^{s_1s_2}\Omega_{s_3;{}^{s_1s_2}g}\Omega_{s_1s_2;g}\\
    &=\omega^{2,1}(s_1s_2,s_3;g)\omega^{2,1}(s_1,s_2;g)X_2,\\
    X_1&=\omega^{2,1}(s_1,s_2s_3;g){}^{s_1}\Omega_{s_2s_3;{}^{s_1}g}\Omega_{s_1;g}\\
    &=\omega^{2,1}(s_1,s_2s_3;g){}^{s_1}\omega^{2,1}(s_2,s_3;{}^{s_1}g)X_2.
    \end{split}
\end{align}
This gives $d_h\omega^{2,1}=0$.
 
\subsection{2+1D}
\subsubsection{Definition of the endpoint operators}
\begin{align}
    U_R(g_1)U_R(g_2)U_R(g_3)&=\Omega_{g_1,g_2}U_R(g_1g_2)U_R(g_3)=\Omega_{g_1,g_2}\Omega_{g_1g_2,g_3}U_R(g_1g_2g_3)\notag\\
    &={}^{g_1}\Omega_{g_2,g_3}U_R(g_1)U_R(g_2g_3)={}^{g_1}\Omega_{g_2,g_3}\Omega_{g_1,g_2g_3}U_R(g_1g_2g_3),\\
    V(s)U_{s^{-1}R}({}^sg_1)U_{s^{-1}R}({}^sg_2)&=\Omega_{s;g_1}U_R(g_1)V(s)U_{s^{-1}R}({}^sg_2)=\Omega_{s;g_1}{}^{g_1}\Omega_{s;g_2}U_R(g_1)U_R(g_2)V(s)\notag\\
    &=\Omega_{s;g_1}{}^{g_1}\Omega_{s;g_2}\Omega_{g_1,g_2}U_R(g_1g_2)V(s)\notag\\
    &={}^s\Omega_{{}^sg_1,{}^sg_2}V(s)U_{s^{-1}R}({}^s(g_1g_2))={}^s\Omega_{{}^sg_1,{}^sg_2}\Omega_{s;g_1g_2}U_R(g_1g_2)V(s),\\
    V(s_1)V(s_2)U_{(s_1s_2)^{-1}R}({}^{s_1s_2}g)&=\Omega_{s_1s_2;g}U_R(g)V(s_1s_2)\notag\\
    &={}^{s_1}\Omega_{s_2;{}^{s_1}g}V(s_1)U_{s_1^{-1}R}({}^{s_1}g)V(s_2)={}^{s_1}\Omega_{s_2;{}^{s_1}g}\Omega_{s_1;g}U_R(g)V(s_1s_2).
\end{align}
Comparing the two evaluations gives \eqref{2D 0-form (0,3)}-\eqref{2D 0-form (2,1)}. The extended boundary operators $\hat{\Omega}$'s with a trivial GNVW index satisfy the same condition. By taking a truncation $I\subset\partial R$, the definition of the endpoint operators $\Gamma$'s supported near $\partial I$ are given by
\begin{align}
    \begin{split}
        \Omega_{g_1,g_2}\Omega_{g_1g_2,g_3}&:=\Gamma_{g_1,g_2,g_3}{}^{g_1}\Omega_{g_2,g_3}\Omega_{g_1,g_2g_3},\\
        \Omega_{s;g_1}{}^{g_1}\Omega_{s;g_2}\Omega_{g_1,g_2}&:=\Gamma_{s;g_1,g_2}{}^s\Omega_{{}^sg_1,{}^sg_2}\Omega_{s;g_1g_2},\\
        \Omega_{s_1s_2;g}&:=\Gamma_{s_1,s_2;g}{}^{s_1}\Omega_{s_2;{}^{s_1}g}\Omega_{s_1;g}.
    \end{split}
\end{align}

\subsubsection{Definition of the indices}
For an operator $\mathcal{O}$ supported near $I$, follow from $U_I(g_1)U_I(g_2)=\Omega_{g_1,g_2}U_I(g_1g_2)$ and $V(s)U_{s^{-1}I}({}^sg)=\Omega_{s;g}U_I(g)V(s)$, the operators
\begin{align}
    &{}^{(g_1,g_2)}\left({}^{g_1g_2}\mathcal{O}\right){}^{g_1}({}^{g_2}\mathcal{O})^{-1},\quad{}^{(s;g)}\left({}^{g}({}^{s}\mathcal{O})\right){}^{s}({}^{{}^sg}\mathcal{O})^{-1}
\end{align}
are supported near $\partial I$. Hence, we define $\Delta$'s supported near $\partial I$.
\begin{align}
    \begin{split}
    \Omega_{g_1,g_2}{}^{g_1g_2}\Omega_{-}&:=\Delta_{g_1,g_2,(-)}{}^{g_1}({}^{g_2}\Omega_{-})\Omega_{g_1,g_2},\\
    \Omega_{g_1;s}{}^{g_1}({}^{s}\Omega_{-})&:=\Delta_{g_1,s,(-)}{}^{s}({}^{{}^s g_1}\Omega_{-})\Omega_{g_1;s}.
    \end{split}
\end{align}
Here we also suppress the subscript that indicates the support of each $\Delta$.

\paragraph{(0,4)} Take $X_1=\Omega_{g_1,g_2}\Omega_{g_1g_2,g_3}\Omega_{g_1g_2g_3,g_4}$ and $X_2:={}^{g_1}({}^{g_2}\Omega_{g_3,g_4}){}^{g_1}\Omega_{g_2,g_3g_4}\Omega_{g_1,g_2g_3g_4}$.
\begin{align}
    \begin{split}
    X_1&=\Gamma_{g_1,g_2,g_3}{}^{g_1}\Omega_{g_2,g_3}\Omega_{g_1,g_2g_3}\Omega_{g_1g_2g_3,g_4}\\
    &=\Gamma_{g_1,g_2,g_3}{}^{{}^{g_1}(g_2,g_3)}\Gamma_{g_1,g_2g_3,g_4}{}^{g_1}\left(\Omega_{g_2,g_3}\Omega_{g_2g_3,g_4}\right)\Omega_{g_1,g_2g_3g_4}\\
    &=\Gamma_{g_1,g_2,g_3}{}^{{}^{g_1}(g_2,g_3)}\Gamma_{g_1,g_2g_3,g_4}{}^{g_1}\Gamma_{g_2,g_3,g_4}X_2,\\
    X_1&={}^{(g_1,g_2)}\Gamma_{g_1g_2,g_3,g_4}\Omega_{g_1,g_2}{}^{g_1g_2}\Omega_{g_3,g_4}\Omega_{g_1g_2,g_3g_4}\\
    &={}^{(g_1,g_2)}\Gamma_{g_1g_2,g_3,g_4}\Delta_{g_1,g_2,(g_3,g_4)}{}^{g_1}({}^{g_2}\Omega_{g_3,g_4})\Omega_{g_1,g_2}\Omega_{g_1g_2,g_3g_4}\\
    &={}^{(g_1,g_2)}\Gamma_{g_1g_2,g_3,g_4}\Delta_{g_1,g_2,(g_3,g_4)}{}^{{}^{g_1\cdot g_2}(g_3,g_4)}\Gamma_{g_1,g_2,g_3g_4}X_2.
    \end{split}
\end{align}
 
\paragraph{(1,3)} Take $X_1=\Omega_{s;g_1}{}^{g_1}\Omega_{s;g_2}\Omega_{g_1,g_2}{}^{g_1g_2}\Omega_{s;g_3}\Omega_{g_1g_2,g_3}$ and $X_2:={}^s({}^{{}^sg_1}\Omega_{{}^sg_2,{}^sg_3}){}^s\Omega_{{}^sg_1,{}^s(g_2g_3)}\Omega_{s;g_1g_2g_3}$.
\begin{align}
    \begin{split}
    X_1&=\Gamma_{s;g_1,g_2}{}^s\Omega_{{}^sg_1,{}^sg_2}\Omega_{s;g_1g_2}{}^{g_1g_2}\Omega_{s;g_3}\Omega_{g_1g_2,g_3}\\
    &=\Gamma_{s;g_1,g_2}{}^{{}^{s}({}^sg_1,{}^sg_2)}\Gamma_{s;g_1g_2,g_3}{}^s\left(\Omega_{{}^sg_1,{}^sg_2}\Omega_{{}^s(g_1g_2),{}^sg_3}\right)\Omega_{s;g_1g_2g_3}\\
    &=\Gamma_{s;g_1,g_2}{}^{{}^{s}({}^sg_1,{}^sg_2)}\Gamma_{s;g_1g_2,g_3}{}^s\Gamma_{{}^sg_1,{}^sg_2,{}^sg_3}X_2,\\
    X_1&={}^{(s;g_1)}\left({}^{{}^{g_1}(s;g_2)}\Delta_{g_1,g_2,(s;g_3)}\right)\Omega_{s;g_1}{}^{g_1}\Omega_{s;g_2}{}^{g_1}({}^{g_2}\Omega_{s;g_3})\Omega_{g_1,g_2}\Omega_{g_1g_2,g_3}\\
    &={}^{(s;g_1)}\left({}^{{}^{g_1}(s;g_2)}\Delta_{g_1,g_2,(s;g_3)}\right){}^{(s;g_1)}\left({}^{{}^{g_1}(s;g_2)}\left({}^{{}^{g_1\cdot g_2}(s;g_3)}\Gamma_{g_1,g_2,g_3}\right)\right)\Omega_{s;g_1}{}^{g_1}\left(\Omega_{s;g_2}{}^{g_2}\Omega_{s;g_3}\Omega_{g_2,g_3}\right)\Omega_{g_1,g_2g_3}\\
    &={}^{(s;g_1)}\left({}^{{}^{g_1}(s;g_2)}\Delta_{g_1,g_2,(s;g_3)}\right){}^{(s;g_1)}\left({}^{{}^{g_1}(s;g_2)}\left({}^{{}^{g_1\cdot g_2}(s;g_3)}\Gamma_{g_1,g_2,g_3}\right)\right)\\
    &\qquad\qquad\qquad{}^{(s;g_1)}({}^{g_1}\Gamma_{s;g_2,g_3})
    \Omega_{s;g_1}{}^{g_1}({}^s\Omega_{{}^sg_2,{}^sg_3}){}^{g_1}\Omega_{s;g_2g_3}\Omega_{g_1,g_2g_3}\\
    &={}^{(s;g_1)}\left({}^{{}^{g_1}(s;g_2)}\Delta_{g_1,g_2,(s;g_3)}\right){}^{(s;g_1)}\left({}^{{}^{g_1}(s;g_2)}\left({}^{{}^{g_1\cdot g_2}(s;g_3)}\Gamma_{g_1,g_2,g_3}\right)\right)\\
    &\qquad\qquad\qquad{}^{(s;g_1)}({}^{g_1}\Gamma_{s;g_2,g_3})\Delta_{s,g_1,({}^sg_2,{}^sg_3)}{}^s({}^{{}^sg_1}\Omega_{{}^sg_2,{}^sg_3})\Omega_{s;g_1}{}^{g_1}\Omega_{s;g_2g_3}\Omega_{g_1,g_2g_3}\\
    &={}^{(s;g_1)}\left({}^{{}^{g_1}(s;g_2)}\Delta_{g_1,g_2,(s;g_3)}\right){}^{(s;g_1)}\left({}^{{}^{g_1}(s;g_2)}\left({}^{{}^{g_1\cdot g_2}(s;g_3)}\Gamma_{g_1,g_2,g_3}\right)\right)\\
    &\qquad\qquad\qquad{}^{(s;g_1)}({}^{g_1}\Gamma_{s;g_2,g_3})\Delta_{s,g_1,({}^sg_2,{}^sg_3)}{}^{{}^{s\cdot{}^{s}g_1}({}^sg_2,{}^sg_3)}\Gamma_{s;g_1,g_2g_3}X_2.
    \end{split}
\end{align}
 
\paragraph{(2,2)} Take $X_1=\Omega_{s_1s_2;g_1}{}^{g_1}\Omega_{s_1s_2;g_2}\Omega_{g_1,g_2}$ and $X_2:={}^{s_1s_2}\Omega_{{}^{s_1s_2}g_1,{}^{s_1s_2}g_2}{}^{s_1}\Omega_{s_2;{}^{s_1}(g_1g_2)}\Omega_{s_1;g_1g_2}$.
\begin{align}
    \begin{split}
    X_1&=\Gamma_{s_1s_2;g_1,g_2}{}^{s_1s_2}\Omega_{{}^{s_1s_2}g_1,{}^{s_1s_2}g_2}\Omega_{s_1s_2;g_1g_2}\\
    &=\Gamma_{s_1s_2;g_1,g_2}{}^{{}^{s_1s_2}({}^{s_1s_2}g_1,{}^{s_1s_2}g_2)}\Gamma_{s_1,s_2;g_1g_2}X_2,\\
    X_1&=\Gamma_{s_1,s_2;g_1}{}^{{}^{s_1}(s_2;{}^{s_1}g_1)}\left({}^{(s_1;g_1)}\left({}^{g_1}\Gamma_{s_1,s_2;g_2}\right)\right){}^{s_1}\Omega_{s_2;{}^{s_1}g_1}\Omega_{s_1;g_1}{}^{g_1}({}^{s_1}\Omega_{s_2;{}^{s_1}g_2}){}^{g_1}\Omega_{s_1;g_2}\Omega_{g_1,g_2}\\
    &=\Gamma_{s_1,s_2;g_1}{}^{{}^{s_1}(s_2;{}^{s_1}g_1)}\left({}^{(s_1;g_1)}\left({}^{g_1}\Gamma_{s_1,s_2;g_2}\right)\right){}^{{}^{s_1}(s_2;{}^{s_1}g_1)}\Delta_{s_1,g_1,(s_2;{}^{s_1}g_1)}{}^{s_1}\Omega_{s_2;{}^{s_1}g_1}{}^{s_1}({}^{{}^{s_1}g_1}\Omega_{s_2;{}^{s_1}g_2})\Omega_{s_1;g_1}{}^{g_1}\Omega_{s_1;g_2}\Omega_{g_1,g_2}\\
    &=\Gamma_{s_1,s_2;g_1}{}^{{}^{s_1}(s_2;{}^{s_1}g_1)}\left({}^{(s_1;g_1)}\left({}^{g_1}\Gamma_{s_1,s_2;g_2}\right)\right){}^{{}^{s_1}(s_2;{}^{s_1}g_1)}\Delta_{s_1,g_1,(s_2;{}^{s_1}g_1)}{}^{{}^{s_1}(s_2;{}^{s_1}g_1)}\left({}^{{}^{s_1\cdot{}^{s_1}g_1}(s_2;{}^{s_1}g_2)}\Gamma_{s_1;g_1,g_2}\right)\\
    &\qquad\qquad\qquad{}^{s_1}\left(\Omega_{s_2;{}^{s_1}g_1}{}^{{}^{s_1}g_1}\Omega_{s_2;{}^{s_1}g_2}\Omega_{{}^{s_1}g_1,{}^{s_1}g_2}\right)\Omega_{s_1;g_1g_2}\\
    &=\Gamma_{s_1,s_2;g_1}{}^{{}^{s_1}(s_2;{}^{s_1}g_1)}\left({}^{(s_1;g_1)}\left({}^{g_1}\Gamma_{s_1,s_2;g_2}\right)\right){}^{{}^{s_1}(s_2;{}^{s_1}g_1)}\Delta_{s_1,g_1,(s_2;{}^{s_1}g_1)}{}^{{}^{s_1}(s_2;{}^{s_1}g_1)}\left({}^{{}^{s_1\cdot{}^{s_1}g_1}(s_2;{}^{s_1}g_2)}\Gamma_{s_1;g_1,g_2}\right)\\
    &\qquad\qquad\qquad{}^{s_1}\Gamma_{s_2;{}^{s_1}g_1,{}^{s_1}g_2}X_2.
    \end{split}
\end{align}
 
\paragraph{(3,1)} Take $X_1=\Omega_{s_1s_2s_3;g}$ and $X_2:={}^{s_1s_2}\Omega_{s_3;{}^{s_1s_2}g}{}^{s_1}\Omega_{s_2;{}^{s_1}g}\Omega_{s_1;g}$.
\begin{align}
    \begin{split}
    X_1&=\Gamma_{s_1,s_2s_3;g}{}^{s_1}\Omega_{s_2s_3;{}^{s_1}g}\Omega_{s_1;g}\\
    &=\Gamma_{s_1,s_2s_3;g}{}^{s_1}\Gamma_{s_2,s_3;{}^{s_1}g}X_2,\\
    X_1&=\Gamma_{s_1s_2,s_3;g}{}^{s_1s_2}\Omega_{s_3;{}^{s_1s_2}g}\Omega_{s_1s_2;g}\\
    &=\Gamma_{s_1s_2,s_3;g}{}^{{}^{s_1s_2}(s_3;{}^{s_1s_2}g)}\Gamma_{s_1,s_2;g}X_2.
    \end{split}
\end{align}

Comparing the two evaluations gives \eqref{2D 0-form (0,4)}-\eqref{2D 0-form (3,1)}. Keep only the operators supported near the right endpoint $I^+$. The definition of $\omega$'s are given by
\begin{align}
    \begin{split}
    &\Gamma_{g_1,g_2,g_3}{}^{{}^{g_1}(g_2,g_3)}\Gamma_{g_1,g_2g_3,g_4}{}^{g_1}\Gamma_{g_2,g_3,g_4}\\
    :=&\omega^{0,4}(g_1,g_2,g_3,g_4){}^{(g_1,g_2)}\Gamma_{g_1g_2,g_3,g_4}\Delta_{g_1,g_2,(g_3,g_4)}{}^{{}^{g_1\cdot g_2}(g_3,g_4)}\Gamma_{g_1,g_2,g_3g_4},\\
    &\Gamma_{s;g_1,g_2}{}^{{}^{s}({}^sg_1,{}^sg_2)}\Gamma_{s;g_1g_2,g_3}{}^s\Gamma_{{}^sg_1,{}^sg_2,{}^sg_3}\\
    :=&\omega^{3,1}(s;g_1,g_2,g_3){}^{(s;g_1)}\left({}^{{}^{g_1}(s;g_2)}\Delta_{g_1,g_2,(s;g_3)}\right){}^{(s;g_1)}\left({}^{{}^{g_1}(s;g_2)}\left({}^{{}^{g_1\cdot g_2}(s;g_3)}\Gamma_{g_1,g_2,g_3}\right)\right)\\
    &\qquad{}^{(s;g_1)}({}^{g_1}\Gamma_{s;g_2,g_3})\Delta_{s,g_1,({}^sg_2,{}^sg_3)}{}^{{}^{s\cdot{}^{s}g_1}({}^sg_2,{}^sg_3)}\Gamma_{s;g_1,g_2g_3},\\
    &\Gamma_{s_1,s_2;g_1}{}^{{}^{s_1}(s_2;{}^{s_1}g_1)}\left({}^{(s_1;g_1)}({}^{g_1}\Gamma_{s_1,s_2;g_2})\Delta_{s_1,g_1,(s_2;{}^{s_1}g_2)}{}^{{}^{s_1\cdot{}^{s_1}g_1}(s_2;{}^{s_1}g_2)}\Gamma_{s_1;g_1,g_2}\right){}^{s_1}\Gamma_{s_2;{}^{s_1}g_1,{}^{s_1}g_2}\\
    :=&\omega^{2,2}(s_1,s_2;g_1,g_2)\Gamma_{s_1s_2;g_1,g_2}{}^{{}^{s_1s_2}({}^{s_1s_2}g_1,{}^{s_1s_2}g_2)}\Gamma_{s_1,s_2;g_1g_2},\\
    &\Gamma_{s_1,s_2s_3;g}{}^{s_1}\Gamma_{s_2,s_3;{}^{s_1}g}\\
    :=&\omega^{3,1}(s_1,s_2,s_3;g)\Gamma_{s_1s_2,s_3;g}{}^{{}^{s_1s_2}(s_3;{}^{s_1s_2}g)}\Gamma_{s_1,s_2;g}.
    \end{split}
\end{align}

\subsubsection{The cocycle conditions}
We first introduce some relations to derive the cocycle conditions.

Since $\Gamma_{g_1,g_2,g_3}{}^{g_1}\Omega_{g_2,g_3}\Omega_{g_1,g_2g_3}=\Omega_{g_1,g_2}\Omega_{g_1g_2,g_3}$, we obtain
\begin{align*}
    &\Gamma_{g_1,g_2,g_3}{}^{{}^{g_1}(g_2,g_3)}\Delta_{g_1,g_2g_3,(-)}{}^{g_1}\Delta_{g_2,g_3,(-)}\\
    =&\Gamma_{g_1,g_2,g_3}{}^{g_1}\Omega_{g_2,g_3}\Omega_{g_1,g_2g_3}{}^{g_1g_2g_3}\Omega_{-}\Omega_{g_1,g_2g_3}^{-1}{}^{g_1}\Omega_{g_2,g_3}^{-1}{}^{g_1\cdot g_2\cdot g_3}\Omega_{-}^{-1}\\
    =&\Omega_{g_1,g_2}\Omega_{g_1g_2,g_3}{}^{g_1g_2g_3}\Omega_{-}\Omega_{g_1g_2,g_3}^{-1}\Omega_{g_1,g_2}^{-1}\Gamma_{g_1,g_2,g_3}{}^{g_1\cdot g_2\cdot g_3}\Omega_{-}^{-1}\\
    =&{}^{(g_1,g_2)}\Delta_{g_1g_2,g_3,(-)}\Delta_{g_1,g_2,{}^{g_3}(-)}{}^{{}^{g_1\cdot g_2\cdot g_3}(-)}\Gamma_{g_1,g_2,g_3},\tag{R1}\\
    &\Delta_{g_1,g_2,(g_3,g_4)}{}^{{}^{g_1\cdot g_2}(g_3,g_4)}\Delta_{g_1,g_2,(g_3g_4,g_5)}{}^{g_1\cdot g_2}\Gamma_{g_3,g_4,g_5}\\
    =&{}^{(g_1,g_2)}({}^{g_1g_2}(\Omega_{g_3,g_4}\Omega_{g_3g_4,g_5})){}^{g_1\cdot g_2}(\Omega_{g_3g_4,g_5}^{-1}\Omega_{g_3,g_4}^{-1}\Gamma_{g_3,g_4,g_5})\\
    =&{}^{(g_1,g_2)}({}^{g_1g_2}(\Gamma_{g_3,g_4,g_5}{}^{g_3}\Omega_{g_4,g_5}\Omega_{g_3,g_4g_5})){}^{g_1\cdot g_2}(\Omega_{g_3,g_4g_5}^{-1}{}^{g_3}\Omega_{g_4,g_5}^{-1})\\
    =&{}^{(g_1,g_2)}({}^{g_1g_2}\Gamma_{g_3,g_4,g_5})\Delta_{g_1,g_2,{}^{g_3}(g_4,g_5)}{}^{{}^{g_1\cdot g_2\cdot g_3}(g_4,g_5)}\Delta_{g_1,g_2,(g_3,g_4g_5)},\tag{R2}\\
    &\Delta_{s,g_1,({}^{s}g_2,{}^{s}g_3)}{}^{{}^{s\cdot {}^{s}g_1}({}^{s}g_2,{}^{s}g_3)}\Delta_{s,g_1,({}^{s}(g_2g_3),{}^{s}g_4)}{}^{s\cdot {}^{s}g_1}\Gamma_{{}^{s}g_2,{}^{s}g_3,{}^{s}g_4}\\
    =&{}^{(s;g_1)}({}^{g_1\cdot s}(\Omega_{{}^{s}g_2,{}^{s}g_3}\Omega_{{}^{s}(g_2g_3),{}^{s}g_4})){}^{s\cdot {}^{s}g_1}(\Omega_{{}^{s}(g_2g_3),{}^{s}g_4}^{-1}\Omega_{{}^{s}g_2,{}^{s}g_3}^{-1}\Gamma_{{}^{s}g_2,{}^{s}g_3,{}^{s}g_4})\\
    =&{}^{(s;g_1)}({}^{g_1\cdot s}(\Gamma_{{}^{s}g_2,{}^{s}g_3,{}^{s}g_4}{}^{{}^{s}g_2}\Omega_{{}^{s}g_3,{}^{s}g_4}\Omega_{{}^{s}g_2,{}^{s}g_3{}^{s}g_4})){}^{s\cdot {}^{s}g_1}(\Omega_{{}^{s}g_2,{}^{s}g_3{}^{s}g_4}^{-1}{}^{{}^{s}g_2}\Omega_{{}^{s}g_3,{}^{s}g_4}^{-1})\\
    =&{}^{(s,g_1)}({}^{g_1\cdot s}\Gamma_{{}^{s}g_2,{}^{s}g_3,{}^{s}g_4})\Delta_{s,g_1,{}^{{}^{s}g_2}({}^{s}g_3,{}^{s}g_4)}{}^{{}^{s\cdot {}^{s}g_1\cdot {}^{s}g_2}({}^{s}g_3,{}^{s}g_4)}\Delta_{s,g_1,({}^{s}g_2,{}^{s}(g_3g_4))}.\tag{R3}
\end{align*}

Since $\Gamma_{s;g_1,g_2}{}^{s}\Omega_{{}^{s}g_1,{}^{s}g_2}\Omega_{s;g_1g_2}=\Omega_{s;g_1}{}^{g_1}\Omega_{s;g_2}\Omega_{g_1,g_2}$, we obtain
\begin{align*}
    &\Gamma_{s;g_1,g_2}{}^{{}^{s}({}^{s}g_1,{}^{s}g_2)}\Delta_{s,g_1g_2,(-)}{}^{s}\Delta_{{}^{s}g_1,{}^{s}g_2,(-)}\\
    =&\Gamma_{s;g_1,g_2}{}^{s}\Omega_{{}^{s}g_1,{}^{s}g_2}\Omega_{s;g_1g_2}{}^{g_1g_2\cdot s}\Omega_{-}\Omega_{s;g_1g_2}^{-1}{}^{s}\Omega_{{}^{s}g_1,{}^{s}g_2}^{-1}{}^{s\cdot {}^{s}g_1\cdot {}^{s}g_2}\Omega_{-}^{-1}\\
    =&\Omega_{s;g_1}{}^{g_1}\Omega_{s;g_2}\Omega_{g_1,g_2}{}^{g_1g_2\cdot s}\Omega_{-}\Omega_{g_1,g_2}^{-1}{}^{g_1}\Omega_{s;g_2}^{-1}\Omega_{s;g_1}^{-1}\Gamma_{s;g_1,g_2}{}^{s\cdot {}^{s}g_1\cdot {}^{s}g_2}\Omega_{-}^{-1}\\
    =&{}^{(s;g_1)}\left({}^{{}^{g_1}(s;g_2)}\Delta_{g_1,g_2,{}^{s}(-)}{}^{g_1}\Delta_{s,g_2,(-)}\right)\Delta_{s,g_1,{}^{{}^{s}g_2}(-)}{}^{{}^{s\cdot {}^{s}g_1\cdot {}^{s}g_2}(-)}\Gamma_{s;g_1,g_2},\tag{R4}\\
    &\Delta_{g_1,g_2,(s;g_3)}{}^{{}^{g_1\cdot g_2}(s;g_3)}\left(\Delta_{g_1,g_2,{}^{g_3}(s;g_4)}{}^{{}^{g_1\cdot g_2\cdot g_3}(s;g_4)}\Delta_{g_1,g_2,(g_3,g_4)}\right){}^{g_1\cdot g_2}\Gamma_{s;g_3,g_4}\\
    =&{}^{(g_1,g_2)}({}^{g_1g_2}(\Omega_{s;g_3}{}^{g_3}\Omega_{s;g_4}\Omega_{g_3,g_4})){}^{g_1\cdot g_2}(\Omega_{g_3,g_4}^{-1}{}^{g_3}\Omega_{s;g_4}^{-1}\Omega_{s;g_3}^{-1}\Gamma_{s;g_3,g_4})\\
    =&{}^{(g_1,g_2)}({}^{g_1g_2}(\Gamma_{s;g_3,g_4}{}^{s}\Omega_{{}^{s}g_3,{}^{s}g_4}\Omega_{s;g_3g_4})){}^{g_1\cdot g_2}(\Omega_{s;g_3g_4}^{-1}{}^{s}\Omega_{{}^{s}g_3,{}^{s}g_4}^{-1})\\
    =&{}^{(g_1,g_2)}({}^{g_1g_2}\Gamma_{s;g_3,g_4})\Delta_{g_1,g_2,{}^{s}({}^{s}g_3,{}^{s}g_4)}{}^{{}^{g_1\cdot g_2\cdot s}({}^{s}g_3,{}^{s}g_4)}\Delta_{g_1,g_2,(s;g_3g_4)},\tag{R5}\\
    &\Delta_{s_1,g_1,(s_2,{}^{s_1}g_2)}{}^{{}^{s_1\cdot {}^{s_1}g_1}(s_2,{}^{s_1}g_2)}\left(\Delta_{s_1,g_1,{}^{{}^{s_1}g_2}(s_2,{}^{s_1}g_3)}{}^{{}^{s_1\cdot {}^{s_1}g_1\cdot {}^{s_1}g_2}(s_2;{}^{s_1}g_3)}\Delta_{s_1,g_1,({}^{s_1}g_2,{}^{s_1}g_3)}\right){}^{s_1\cdot {}^{s_1}g_1}\Gamma_{s_2;{}^{s_1}g_2,{}^{s_1}g_3}\\
    =&{}^{(s_1;g_1)}({}^{g_1\cdot s_1}(\Omega_{s_2;{}^{s_1}g_2}
    {}^{{}^{s_1}g_2}\Omega_{s_2;{}^{s_1}g_3}\Omega_{{}^{s_1}g_2,{}^{s_1}g_3})){}^{s_1\cdot {}^{s_1}g_1}(\Omega_{{}^{s_1}g_2,{}^{s_1}g_3}^{-1}{}^{{}^{s_1}g_2}\Omega_{s_2;{}^{s_1}g_3}^{-1}\Omega_{s_2;{}^{s_1}g_2}^{-1}\Gamma_{s_2;{}^{s_1}g_2,{}^{s_1}g_3})\\
    =&{}^{(s_1;g_1)}({}^{g_1\cdot s_1}(\Gamma_{s_2;{}^{s_1}g_2,{}^{s_1}g_3}{}^{s_2}\Omega_{{}^{s_1s_2}g_2,{}^{s_1s_2}g_3}\Omega_{s_2;{}^{s_1}(g_2g_3)})){}^{s_1\cdot {}^{s_1}g_1}(\Omega_{s_2;{}^{s_1}(g_2g_3)}^{-1}{}^{s_2}\Omega_{{}^{s_1s_2}g_2,{}^{s_1s_2}g_3}^{-1})\\
    =&{}^{(s_1;g_1)}({}^{g_1\cdot s_1}\Gamma_{s_2;{}^{s_1}g_2,{}^{s_1}g_3})\Delta_{s_1,g_1,{}^{s_2}({}^{s_1s_2}g_2,{}^{s_1s_2}g_3)}{}^{{}^{s\cdot {}^{s_1}g_1\cdot s_2}({}^{s_1s_2}g_2,{}^{s_1s_2}g_3)}\Delta_{s_1,g_1,(s_2;{}^{s_1}(g_2g_3))}.\tag{R6}
\end{align*}

Since $\Gamma_{s_1,s_2;g}{}^{s_1}\Omega_{s_2;{}^{s_1}g}\Omega_{s_1;g}=\Omega_{s_1s_2,g}$, we obtain
\begin{align*}
    &\Gamma_{s_1,s_2;g}{}^{{}^{s_1}(s_2;{}^{s_1}g)}\Delta_{s_1,g,{}^{s_2}(-)}{}^{s_1}\Delta_{s_2,{}^{s_1}g,(-)}\\
    =&\Gamma_{s_1,s_2;g}{}^{s_1}\Omega_{s_2;{}^{s_1}g}\Omega_{s_1;g}{}^{g\cdot s_1s_2}\Omega_{-}\Omega_{s_1;g}^{-1}{}^{s_1}\Omega_{s_2;{}^{s_1}g}^{-1}{}^{s_1s_2\cdot {}^{s_1s_2}g}\Omega_{-}^{-1}\\
    =&\Omega_{s_1s_2;g}{}^{g\cdot s_1s_2}\Omega_{-}\Omega_{s_1s_2;g}^{-1}\Gamma_{s_1,s_2;g}{}^{s_1s_2\cdot {}^{s_1s_2}g}\Omega_{-}^{-1}\\
    =&\Delta_{s_1s_2,g,(-)}{}^{{}^{s_1s_2\cdot {}^{s_1s_2}g}(-)}\Gamma_{s_1,s_2;g},\tag{R7}\\
    &\Delta_{g_1,g_2,(s_1s_2;g_3)}{}^{g_1\cdot g_2}\Gamma_{s_1,s_2;g_3}\\
    =&{}^{(g_1,g_2)}({}^{g_1g_2}(\Omega_{s_1s_2;g_3})){}^{g_1\cdot g_2}(\Omega_{s_1s_2;g_3}^{-1}\Gamma_{s_1,s_2;g_3})\\
    =&{}^{(g_1,g_2)}({}^{g_1g_2}(\Gamma_{s_1,s_2;g_3}{}^{s_1}\Omega_{s_2;{}^{s_1}g_3}\Omega_{s_1;g_3})){}^{g_1\cdot g_2}(\Omega_{s_1;g_3}^{-1}{}^{s_1}\Omega_{s_2;{}^{s_1}g_3}^{-1})\\
    =&{}^{(g_1,g_2)}({}^{g_1g_2}\Gamma_{s_1,s_2;g_3})\Delta_{g_1,g_2,{}^{s_1}(s_2;{}^{s_1}g_3)}{}^{{}^{g_1\cdot g_2\cdot s_1}(s_2;{}^{s_1}g_3)}\Delta_{g_1,g_2,(s_1;g_3)},\tag{R8}\\
    &\Delta_{s_1,g_1,(s_2s_3;{}^{s_1}g_2)}{}^{s_1\cdot {}^{s_1}g_1}\Gamma_{s_2,s_3;{}^{s_1}g_2}\\
    =&{}^{(s_1;g_1)}({}^{g_1\cdot s_1}(\Omega_{s_2s_3;{}^{s_1}g_2})){}^{s_1\cdot {}^{s_1}g_1}(\Omega_{s_2s_3;{}^{s_1}g_2}^{-1}\Gamma_{s_2,s_3;{}^{s_1}g_2})\\
    =&{}^{(s_1;g_1)}({}^{g_1\cdot s_1}(\Gamma_{s_2,s_3;{}^{s_1}g_2}{}^{s_2}\Omega_{s_3;{}^{s_1s_2}g_2}\Omega_{s_2;{}^{s_1}g_2})){}^{s_1\cdot {}^{s_1}g_1}(\Omega_{s_2;{}^{s_1}g_2}^{-1}{}^{s_2}\Omega_{s_3;{}^{s_1s_2}g_2}^{-1})\\
    =&{}^{(s_1;g_1)}({}^{g_1\cdot s_1}\Gamma_{s_2,s_3;{}^{s_1}g_2})\Delta_{s_1,g_1,{}^{s_2}(s_3;{}^{s_1s_2}g_2)}{}^{{}^{s_1\cdot {}^{s_1}g_1\cdot s_2}(s_3;{}^{s_1s_2}g_2)}\Delta_{s_1,g_1,(s_2;{}^{s_1}g_2)}.\tag{R9}
\end{align*}

The following cocycle conditions can be shown by combinations of the definition of $\omega$'s and relations (R1)--(R9).

\paragraph{(0,5)} Take
\begin{align}
    \begin{split}
        &Y_1=\Gamma_{g_1,g_2,g_3}{}^{{}^{g_1}(g_2,g_3)}\Gamma_{g_1,g_2g_3,g_4}{}^{g_1}\Gamma_{g_2,g_3,g_4}{}^{{}^{g_1\cdot g_2}(g_3,g_4)}\left({}^{{}^{g_1}(g_2,g_3g_4)}\Gamma_{g_1,g_2g_3g_4,g_5}{}^{g_1}\Gamma_{g_2,g_3g_4,g_5}\right){}^{g_1\cdot g_2}\Gamma_{g_3,g_4,g_5},\\
        &Y_2={}^{(g_1,g_2)}\left(
        {}^{(g_1g_2,g_3)}\Gamma_{g_1g_2g_3,g_4,g_5}\Delta_{g_1g_2,g_3,(g_4,g_5)}\right)\Delta_{g_1,g_2,{}^{g_3}(g_4,g_5)}\\
        &\qquad{}^{{}^{g_1\cdot g_2\cdot g_3}(g_4,g_5)}\left({}^{(g_1,g_2)}\Gamma_{g_1g_2,g_3,g_4g_5}\Delta_{g_1,g_2,(g_3,g_4g_5)}{}^{{}^{g_1\cdot g_2}(g_3,g_4g_5)}\Gamma_{g_1,g_2,g_3g_4g_5}\right).
    \end{split}
\end{align}
By the definition of $\omega^{0,4}$ and the relation (R2), we obtain
\begin{equation}
    Y_1=\omega^{0,4}(g_1,g_2,g_3,g_4)\omega^{0,4}(g_1,g_2,g_3g_4,g_5)\omega^{0,4}(g_1g_2,g_3,g_4,g_5)Y_2.
\end{equation}
Meanwhile, by the definition of $\omega^{0,4}$ and the relation (R1), we obtain
\begin{equation}
    Y_1=\omega^{0,4}(g_2,g_3,g_4,g_5)\omega^{0,4}(g_1,g_2g_3,g_4,g_5)\omega^{0,4}(g_1,g_2,g_3,g_4g_5)Y_2.
\end{equation}
This gives $d_v\omega^{0,4}=0$.

\paragraph{(1,4)} Take
\begin{align}
    \begin{split}
        &Y_1=\Gamma_{s;g_1,g_2}{}^{{}^{s}({}^{s}g_1,{}^{s}g_2)}\Gamma_{s;g_1g_2,g_3}{}^s\Gamma_{{}^sg_1,{}^sg_2,{}^sg_3}{}^{{}^{s\cdot{}^sg_1}({}^sg_2,{}^sg_3)}\left({}^{{}^s({}^sg_1,{}^s(g_2g_3))}\Gamma_{s;g_1g_2g_3,g_4}{}^s\Gamma_{{}^sg_1,{}^s(g_2g_3),{}^sg_4}\right){}^{s\cdot {}^sg_1}\Gamma_{{}^sg_2,{}^sg_3,{}^sg_4}\\
        &Y_2={}^{(s;g_1)}\Bigg({}^{{}^{g_1}(s;g_2)}\Bigg({}^{(g_1,g_2)}\Big({}^{{}^{g_1g_2}(s;g_3)}\left(\Delta_{g_1g_2,g_3,(s;g_4)}{}^{{}^{g_1g_2\cdot g_3}(s;g_4)}\Gamma_{g_1g_2,g_3,g_4}\right){}^{g_1g_2}\Gamma_{s;g_3,g_4}\Big)\Delta_{g_1,g_2,{}^{s}({}^{s}g_3,{}^{s}g_4)}\Bigg)\\
        &\qquad\qquad{}^{g_1}\Delta_{s,g_2,({}^{s}g_3,{}^{s}g_4)}\Bigg)\Delta_{s,g_1,{}^{{}^{s}g_2}({}^{s}g_3,{}^{s}g_4)}{}^{{}^{s\cdot {}^{s}g_1\cdot {}^{s}g_2}({}^{s}g_3,{}^{s}g_4)}\Bigg({}^{(s;g_1)}\Big({}^{{}^{g_1}(s;g_2)}\left(\Delta_{g_1,g_2,(s;g_3g_4)}{}^{{}^{g_1\cdot g_2}(s;g_3g_4)}\Gamma_{g_1,g_2,g_3g_4}\right)\\
        &\qquad\qquad{}^{g_1}\Gamma_{s;g_2,g_3g_4}\Big)\Delta_{s_1,g_1,({}^{s}g_2,{}^{s}(g_3g_4))}{}^{{}^{s\cdot {}^{s}g_1}({}^{s}g_2,{}^{s}(g_3g_4))}\Gamma_{s;g_1,g_2g_3g_4}\Bigg).
    \end{split}
\end{align}
By the definition of $\omega^{0,4}$ and $\omega^{1,3}$, and relations (R1), (R3), and (R5), we obtain
\begin{equation}
    Y_1=\omega^{1,3}(s;g_1,g_2,g_3)\omega^{1,3}(s;g_1,g_2g_3,g_4)\omega^{1,3}(s;g_2,g_3,g_4)\omega^{0,4}(g_1,g_2,g_3,g_4)Y_2.
\end{equation}
Meanwhile, by the definition of $\omega^{0,4}$ and $\omega^{1,3}$, and the relation (R4), we obtain
\begin{equation}
    Y_1={}^{s}\omega^{0,4}({}^{s}g_1,{}^{s}g_2,{}^{s}g_3,{}^{s}g_4)\omega^{1,3}(s;g_1g_2,g_3,g_4)\omega^{1,3}(s;g_1,g_2,g_3g_4)Y_2.
\end{equation}
This gives $d_h\omega^{0,4}-d_v\omega^{1,3}=0$.

\paragraph{(2,3)} Take
\begin{align}
    \begin{split}
        &Y_1=\Gamma_{s_1,s_2;g_1}{}^{{}^{s_1}(s_2;{}^{s_1}g_1)}\Big({}^{(s_1;g_1)}({}^{g_1}\Gamma_{s_1,s_2;g_2})\Delta_{s_1,g_1,(s_2;{}^{s_1}g_2)}{}^{{}^{s_1\cdot{}^{s_1}g_1}(s_2;{}^{s_1}g_2)}\Gamma_{s_1;g_1,g_2}\Big){}^{s_1}\Gamma_{s_2;{}^{s_1}g_1,{}^{s_1}g_2}\\
        &\qquad\qquad{}^{{}^{s_1s_2}({}^{s_1s_2}g_1,{}^{s_1s_2}g_2)}\Bigg({}^{{}^{s_1}(s_2;{}^{s_1}(g_1g_2))}\Big({}^{(s_1;g_1g_2)}({}^{g_1g_2}\Gamma_{s_1,s_2;g_3})\Delta_{s_1,g_1g_2,(s_2;{}^{s_1}g_3)}{}^{{}^{s_1\cdot{}^{s_1}(g_1g_2)}(s_2;{}^{s_1}g_3)}\Gamma_{s_1;g_1g_2,g_3}\Big)\\
        &\qquad\qquad{}^{s_1}\Gamma_{s_2;{}^{s_1}(g_1g_2),{}^{s_1}g_3}\Bigg){}^{s_1s_2}\Gamma_{{}^{s_1s_2}g_1,{}^{s_1s_2}g_2,{}^{s_1s_2}g_3}\\
        &Y_2={}^{(s_1s_2;g)}\Big({}^{{}^{g_1}(s_1s_2;g_2)}(\Delta_{g_1,g_2,(s_1s_2;g_3)}{}^{{}^{g_1\cdot g_2}(s_1s_2;g_3)}\Gamma_{g_1,g_2,g_3}){}^{g_1}\Gamma_{s_1s_2;g_2,g_3}\Big)\Delta_{s_1s_2,g_1,({}^{s_1s_2}g_2,{}^{s_1s_2}g_3)}\\
        &\qquad\qquad{}^{{}^{s_1s_2\cdot{}^{s_1s_2}g_1}({}^{s_1s_2}g_2,{}^{s_1s_2}g_3)}(\Gamma_{s_1s_2;g_1,g_2g_3}{}^{{}^{s_1s_2}({}^{s_1s_2}g_1,{}^{s_1s_2}(g_2g_3))}\Gamma_{s_1,s_2;g_1g_2g_3}).
    \end{split}
\end{align}
By the definition of $\omega^{1,3}$ and $\omega^{2,2}$, we obtain
\begin{equation}
    Y_1=\omega^{2,2}(s_1,s_2;g_1,g_2)\omega^{2,2}(s_1,s_2;g_1g_2,g_3)\omega^{1,3}(s_1s_2;g_1,g_2,g_3)Y_2.
\end{equation}
Meanwhile, by the definition of $\omega^{1,3}$ and $\omega^{2,2}$, and relations (R4), (R6), (R7), and (R8), we obtain
\begin{equation}
    Y_1={}^{s_1}\omega^{1,3}(s_2;{}^{s_1}g_1,{}^{s_1}g_2,{}^{s_1}g_3)\omega^{1,3}(s_1;g_1,g_2,g_3)\omega^{2,2}(s_1,s_2;g_2,g_3)\omega^{2,2}(s_1,s_2;g_1,g_2g_3)Y_2.
\end{equation}
This gives $d_h\omega^{1,3}+d_v\omega^{2,2}=0$.

\paragraph{(3,2)} Take
\begin{align}
    \begin{split}
        &Y_1=\Gamma_{s_1,s_2s_3;g_1}{}^{s_1}\Gamma_{s_2,s_3;{}^{s_1}g_2}{}^{{}^{s_1s_2}(s_3;{}^{s_1s_2}g_1)}\Bigg({}^{{}^{s_1}(s_2;{}^{s_1}g_1)}\Bigg({}^{(s_1;g_1)}\Bigg({}^{g_1}\Big(\Gamma_{s_1,s_2s_3;g_2}{}^{s_1}\Gamma_{s_2,s_3;{}^{s_1}g_2}\Big)\Bigg)\Delta_{s_1,g_1,{}^{s_2}(s_3;{}^{s_1s_2}g_2)}\\
        &\qquad\qquad{}^{{}^{s_1\cdot{}^{s_1}g_1\cdot s_2}(s_3;{}^{s_1s_2}g_2)}\Bigg(\Delta_{s_1,g_1,(s_2;{}^{s_1}g_2)}{}^{{}^{s_1\cdot{}^{s_1}g_1}(s_2;{}^{s_1}g_2)}\Gamma_{s_1;g_1,g_2}\Bigg)\Bigg)\Bigg)\\
        &\qquad\qquad{}^{{}^{s_1s_2}(s_3;{}^{s_1s_2}g_1)}\Bigg({}^{s_1}\Delta_{s_2,{}^{s_1}g_1,(s_3;{}^{s_1s_2}g_2)}{}^{{}^{s_1s_2\cdot{}^{s_1s_2}g_1}(s_3;{}^{s_1s_2}g_2)}\Big({}^{s_1}\Gamma_{s_2;{}^{s_1}g_1,{}^{s_1}g_2}\Big)\Bigg){}^{s_1s_2}\Gamma_{s_3;{}^{s_1s_2}g_1,{}^{s_1s_2}g_2},\\
        &Y_2=\Gamma_{s_1s_2s_3;g_1;g_2}{}^{{}^{s_1s_2s_3}({}^{s_1s_2s_3}g_1,{}^{s_1s_2s_3}g_2)}\left(\Gamma_{s_1s_2,s_3;g_1g_2}{}^{{}^{s_1s_2}(s_3;{}^{s_1s_2}(g_1g_2))}\Gamma_{s_1,s_2;g_1g_2}\right).
    \end{split}
\end{align}
By the definition of $\omega^{2,2}$ and $\omega^{3,1}$, and the relation (R7), we obtain
\begin{equation}
    Y_1=\omega^{3,1}(s_1,s_2,s_3;g_1)\omega^{3,1}(s_1,s_2,s_3;g_2)\omega^{2,2}(s_1,s_2;g_1,g_2)\omega^{2,2}(s_1s_2,s_3;g_1,g_2)Y_2.
\end{equation}
Meanwhile, by the definition of $\omega^{2,2}$ and $\omega^{3,1}$, and the relation (R9), we obtain
\begin{equation}
    Y_1={}^{s}\omega^{2,2}(s_2,s_3;{}^{s_1}g_1,{}^{s_1}g_2)\omega^{2,2}(s_1,s_2s_3;g_1,g_2)\omega^{3,1}(s_1,s_2,s_3;g_1g_2)Y_2.
\end{equation}
This gives $d_h\omega^{2,2}-d_v\omega^{3,1}=0$.

\paragraph{(4,1)} Take
\begin{align}
    \begin{split}
        &Y_1=\Gamma_{s_1,s_2s_3s_4;g}{}^{s_1}\Gamma_{s_2,s_3s_4;g}{}^{s_1s_2}\Gamma_{s_3,s_4;{}^{s_1s_2}g},\\
        &Y_2=\Gamma_{s_1s_2s_3,s_4;g}{}^{{}^{s_1s_2s_3}(s_4;{}^{s_1s_2s_3}g)}\left(\Gamma_{s_1s_2,s_3;g}{}^{{}^{s_1s_2}(s_3;g)}\Gamma_{s_1,s_2;g}\right).
    \end{split}
\end{align}
By the definition of $\omega^{3,1}$, we obtain
\begin{align}
    \begin{split}
        Y_1=&\omega^{3,1}(s_1,s_2,s_3s_4;g)\omega^{3,1}(s_1s_2,s_3,s_4;g)Y_2\\
        =&{}^{s_1}\omega^{3,1}(s_2,s_3,s_4;{}^{s_1}g)\omega^{3,1}(s_1,s_2s_3,s_4;g)\omega^{3,1}(s_1,s_2,s_3;g)Y_2.
    \end{split}
\end{align}
This gives $d_h\omega^{3,1}=0$.
\section{Group Cohomology of Semidirect Products}\label{Cohomology of Semidirect Products}
A semidirect product is a split group extension
\begin{equation}
    1\to K\xrightarrow{i} G\xrightarrow{p} H\to1.
\end{equation}
Choose a group homomorphism $s:H\to G$ that splits the above short exact sequence, we define $\phi: H\to\mathrm{Aut}(K)$ as
\begin{equation}
    i(\phi(h)(k))=s(h)i(k)s(h)^{-1}.
\end{equation}
It is well-defined since $p$ sends the right-hand side to the identity of $H$ and $i$ is injective. Denote $G:=K\rtimes_\phi H$ after fixing a choice of such $s:H\to G$.

Given a $G$-module $M$, the Lyndon-Hochschild-Serre construction gives a double complex \cite{Lyndon,HochschildSerre}
\begin{equation}
    C^{p,q}:=C^p(BH,C^q(BK,M)).
\end{equation}
The $K$-action on $M$ is induced from the $i(K)$-action on $M$, and the $H$-action on $C^q(BK,M)$ is induced from the $s(H)$-action on $M$ together with the $H$-action $\phi:H\to\mathrm{Aut}(K)$ on $K$. More explicitly, $C^{p,q}$ consists of functions $f(h_1,\dots,h_p;k_1,\dots,k_q)$ from $H^p\times K^q$ to $M$, and the vertical and horizontal differentials $d_v: C^{p,q}\to C^{p,q+1}$ and $d_h: C^{p,q}\to C^{p+1,q}$ are given by
\begin{align}
    \begin{split}
        d_vf(h_1,\dots,h_p;k_1,\dots,k_{q+1})&={}^{k_1}f(h_1,\dots,h_p;k_2,\dots,k_{q+1})\\
        &+\sum_{j=1}^q(-1)^jf(h_1,\dots,h_p;k_1,\dots,k_jk_{j+1},\dots,k_{q+1})\\
        &+(-1)^{q+1}f(h_1,\dots,h_p;k_1,\dots,k_q),\\
        d_hf(h_1,\dots,h_{p+1};k_1,\dots,k_q)&={}^{h_1}f(h_2,\dots,h_{p+1};\phi(h_1)^{-1}k_1,\dots,\phi(h_1)^{-1}k_q)\\
        &+\sum_{i=1}^p(-1)^if(h_1,\dots,h_ih_{i+1},\dots,h_{p+1};k_1,\dots,k_q)\\
        &+(-1)^{p+1}f(h_1,\dots,h_p;k_1,\dots,k_q).\\
    \end{split}
\end{align}
Here, ${}^km$ denotes the induced action of $k\in K$ on $m\in M$ and ${}^hm$ denotes the induced action of $h\in H$ on $m\in M$. The total complex is defined to be $\mathrm{Tot}^n(C^{\bullet,\bullet}):=\bigoplus_{p+q=n} C^{p,q}$ equipped with the differential $D:=d_h+(-1)^pd_v$, whose total cohomology computes $H^\bullet(BG,M)$.

We now consider the truncated double complex
\begin{equation}
    \tilde{C}^{p,q}:=\begin{cases}
        C^{p,q}&\text{if }q>0,\\
        0&\text{if }q=0.
    \end{cases}
\end{equation}
The total differential $\tilde{D}$ on this complex is induced from $D$. There is a short exact sequence of cochain complexes
\begin{equation}
    0\to \mathrm{Tot}(\tilde{C}^{\bullet,\bullet})\overset{\iota}{\to}\mathrm{Tot}(C^{\bullet,\bullet})\overset{\pi}{\to}(C^{\bullet,0},d_h)\to0,
\end{equation}
which induces a long exact sequence
\begin{equation}
    \cdots\to H^{n-1}(C^{\bullet,0},d_h)\xrightarrow{\delta} H^n(\mathrm{Tot}(\tilde{C}^{\bullet,\bullet}))\xrightarrow{\iota_*}H^n(\mathrm{Tot}(C^{\bullet,\bullet}))\xrightarrow{\pi_*}H^n(C^{\bullet,0},d_h)\to\cdots.
\end{equation}
This can be rewritten as
\begin{equation}
    \cdots\to H^{n-1}(BH,M)\xrightarrow{\delta} H^n(\mathrm{Tot}(\tilde{C}^{\bullet,\bullet}))\xrightarrow{\iota_*}H^n(BG,M)\xrightarrow{\pi_*}H^n(BH,M)\to\cdots.
\end{equation}

Suppose the $K$-action on $M$ is trivial, which corresponds to the unitary symmetry case in this work. The vertical differentials on the $q=0$ row $d_v:C^{p,0}\to C^{p,1}$ vanish. Consequently, the natural inclusion $\sigma:C^{\bullet,0}\to\mathrm{Tot}(C^{\bullet,\bullet})$ is a cochain map satisfying $\sigma\circ \pi=\mathrm{id}_{(C^{\bullet,0},d_h)}$. Thus, the above short exact sequence of cochain complexes splits. It follows that the connecting homomorphisms $\delta$ in the induced long exact sequence vanish. Therefore, we obtain short exact sequences
\begin{equation}
    0\to H^n(\mathrm{Tot}(\tilde{C}^{\bullet,\bullet}))\xrightarrow{\iota_*}H^n(BG,M)\xrightarrow{\pi_*}H^n(BH,M)\to0,
\end{equation}
and
\begin{equation}
    H^n(\mathrm{Tot}(\tilde{C}^{\bullet,\bullet}))\simeq\ker\left(H^n(BG,M)\overset{\pi_*}{\to}H^n(BH,M)\right).
\end{equation}
In other words, $H^n(\mathrm{Tot}(\tilde{C}^{\bullet,\bullet}))$ is isomorphic to the classes in $H^3(BG,M)$ whose restriction to $s(H)\subseteq G$ is trivial.
\section{Differentials in Spectral Sequences}\label{Spectral Sequences}
Given a double complex $E_0^{p,q}$ with horizontal and vertical differentials
\begin{align}
    \begin{split}
        &d_h:E_0^{p,q}\to E_0^{p+1,q},\\
        &d_v:E_0^{p,q}\to E_0^{p,q+1}
    \end{split}
\end{align}
commute with each other, the total differential $D:=d_h+(-1)^pd_v$ satisfies $D^2=0$.

The first and second pages of the spectral sequence are defined by
\begin{align}
    \begin{split}
        &E_1^{p,q}:=H^q(E_0^{p,\bullet},(-1)^pd_v),\\
        &E_2^{p,q}:=H^p(E_1^{\bullet,q},d_h).
    \end{split}
\end{align}
Given a representative $\lambda_0\in E_0^{p,q}$ of a class $[\lambda_0]\in E_2^{p,q}$, then, 
\begin{align}
    \begin{split}
        &d_v\lambda_0=0\quad\text{in }E_0^{p,q+1},\\
        &[d_h\lambda_0]=0\quad\text{in }E_1^{p+1,q}.
    \end{split}
\end{align}
Hence, we can choose $\lambda_1\in E_0^{p+1,q-1}$ such that
\begin{equation}
    d_h\lambda_0+(-1)^{p+1}d_v\lambda_1=0.
\end{equation}
The differential $d_2^{p,q}:E_2^{p,q}\to E_2^{p+2,q-1}$ on the second page is then defined by
\begin{equation}
    d_2^{p,q}[\lambda_0]=[D(\lambda_0+\lambda_1)]=[d_h \lambda_1].
\end{equation}

Iteratively, we define the $r$-th page of the spectral sequence to be
\begin{equation}
    E_r^{p,q}:=\mathrm{ker}d_{r-1}^{p,q}/\mathrm{im}d_{r-1}^{p-(r-1),q+(r-1)-1}
\end{equation}
for $r>2$. Given a representative $\lambda_0\in E_0^{p,q}$ of a class $[\lambda_0]\in E_r^{p,q}$, then, we can find $\lambda_k\in E_0^{p+k,q-k}$ ($k=1,\dots r-1$) such that
\begin{align}
    \begin{split}
        d_v\lambda_0=&0,\\
        d_h\lambda_0+(-1)^{p+1}d_v\lambda_1=&0,\\
        d_h\lambda_1+(-1)^{p+2}d_v\lambda_2=&0,\\
        &\vdots\\
        d_h\lambda_{r-2}+(-1)^{p+r-1}d_v\lambda_{r-1}=&0.
    \end{split}
\end{align}
The differential $d_r^{p,q}:E_r^{p,q}\to E_r^{p+r,q-r+1}$ is then defined by
\begin{equation}
    d_r^{p,q}[\lambda_0]=[D(\lambda_0+\lambda_1+\dots+\lambda_{r-1})]=[d_h \lambda_{r-1}].
\end{equation}

\bibliographystyle{utphys}
\bibliography{bibliography, crystalline_anomalies_refs}
\end{document}